\documentclass[aps,twocolumn,showpacs,preprintnumbers,nofootinbib,prd,superscriptaddress,groupedaddress,10pt]{revtex4-1}

\usepackage{graphicx,amssymb,amsmath,amsthm,amsfonts,epsfig,epsf}
\usepackage[linktocpage]{hyperref}
\usepackage[usenames]{color}
\usepackage{epstopdf}

\usepackage{bm}
\usepackage{dcolumn}
\usepackage[latin1]{inputenc}
\usepackage{latexsym}
\usepackage{rotating}
\usepackage{hyperref}
\usepackage{color}
\usepackage{longtable}

\usepackage{enumerate}
\usepackage{tensor,multirow}
\usepackage{url}

\usepackage{slashed}

\begin{document}
\title{Post-Newtonian spin-tidal couplings for compact binaries}

\author{Tiziano Abdelsalhin}
\email{tiziano.abdelsalhin@roma1.infn.it}
\affiliation{Dipartimento di Fisica ``Sapienza''
Universit\`a di Roma \& Sezione INFN Roma1,
Piazzale Aldo Moro 5, 00185, Roma, Italy}

\author{Leonardo Gualtieri}
\email{leonardo.gualtieri@roma1.infn.it}
\affiliation{Dipartimento di Fisica ``Sapienza''
Universit\`a di Roma \& Sezione INFN Roma1,
Piazzale Aldo Moro 5, 00185, Roma, Italy}

\author{Paolo Pani}
\email{paolo.pani@roma1.infn.it}
\affiliation{Dipartimento di Fisica ``Sapienza''
Universit\`a di Roma \& Sezione INFN Roma1,
Piazzale Aldo Moro 5, 00185, Roma, Italy}

\begin{abstract}
We compute the spin-tidal couplings that affect the dynamics of two orbiting bodies at the leading order in the
post-Newtonian~(PN) framework and to linear order in the spin. These corrections belong to two classes: (i) terms arising
from the coupling between the ordinary tidal terms and the point-particle terms, which depend on the standard tidal Love
numbers of order $l$ and affect the gravitational-wave~(GW) phase at $(2l+5/2)$PN order and (ii) terms depending on the
rotational tidal Love numbers, recently introduced in previous work, that affect the GW phase at
$(2l+1/2+\delta_{2l})$PN order. For circular orbits and spins orthogonal to the orbital plane, all leading-order
spin-tidal terms enter the GW phase at $1.5$PN order relative to the standard, quadrupolar, tidal deformability term
(and, thus, \emph{before} the standard octupolar tidal deformability terms).  We present the GW phase that includes
all tidal terms up to $6.5$PN order and to linear order in the spin.  We comment on a conceptual issue related to
the inclusion of the rotational tidal Love numbers in a Lagrangian formulation and on the relevance of spin-tidal
couplings for parameter estimation in coalescing neutron-star binaries and for tests of gravity.
\end{abstract}

\maketitle

\section{Introduction}\label{sec:intro}
\subsection{Background and motivation}
GW170817~\cite{TheLIGOScientific:2017qsa} --~the first coalescence of a binary neutron-star (NS) system detected by the
gravitational-wave (GW) interferometers LIGO and Virgo~-- is a milestone in GW astronomy.  With more NS-NS coalescence
signals expected in the near future, it will be possible to constrain the equation of state of the NS
core~\cite{DelPozzo:2013ala,TheLIGOScientific:2017qsa,Bauswein:2017vtn,Most:2018hfd,Harry:2018hke,Annala:2017llu,EoS-GW170817},
to test gravity in the highly relativistic/strong-curvature/supranuclear-density
regime~\cite{TheLIGOScientific:2017qsa}, and to detect coincident electromagnetic signals emitted by these sources in
various bands~\cite{GBM:2017lvd,Monitor:2017mdv}.

A major challenge in the parameter estimation of NS binaries is the modeling of the GW signal during the late inspiral,
merger, and postmerger phases~\cite{Arun:2008kb}. This is typically achieved by using GW templates obtained either
phenomenologically or using the effective-one-body
approach~\cite{Buonanno:1998gg,Bernuzzi:2014owa,Bernuzzi:2015rla,Hinderer:2016eia}, fitted to numerical-relativity
waveforms~\cite{Dietrich:2017feu,Dietrich:2018uni}. A ubiquitous ingredient of these templates is an accurate
description of the early-inspiral phase as described by the post-Newtonian~(PN)
formalism~\cite{Arun:2008kb,Buonanno:2009zt,Mishra:2016whh} (i.e., a weak-field/slow-velocity expansion of Einstein's
equations), where the dynamics of the binary is driven by energy and angular momentum loss, and the two bodies are
modeled as two point particles endowed with a series of multipole moments and with finite-size tidal
corrections~\cite{Blanchet:2006zz,Vines:2011ud,Damour:2012yf}. The latter are encoded in the way a NS responds when
acted upon by the external gravitational field of its companion~-- through the tidal Love numbers (TLNs) (see,
e.g.,~\cite{PoissonWill} and references therein).

To the leading order in the tidal field, the TLNs are proportional to the induced multipole moments. As such, they can
be divided into two categories: \emph{electric} (or even parity) TLNs, which are related to the induced mass multipole
moments, and \emph{magnetic} (or odd parity) TLNs, which are related to the induced current multipole moments and do not
have an analog in Newtonian theory.  Tidal deformability introduces a $5$PN correction to the GW phase relative to the
leading-order GW term~\cite{Flanagan:2007ix,Hinderer:2007mb}, this correction being proportional to the quadrupolar
electric TLN. The next-to-leading order correction from quadrupolar electric TLNs was computed in
Ref.~\cite{Vines:2011ud} and enters at $6$PN order, which is also the leading-order correction\footnote{We warn the
  reader that the quadrupolar magnetic Love numbers affect the GW phase at $6$PN order also for equal-mass binaries, see
  erratum in Ref.~\cite{Yagi:2013sva} and Ref.~\cite{Banihashemi:2018xfb}. This point will be important for the
  following discussion.} from quadrupolar magnetic TLNs~\cite{Yagi:2013sva,Banihashemi:2018xfb}.  Moreover, the leading
tail contribution from quadrupolar electric TLNs, appearing at $6.5$PN order, has been computed in
Ref.~\cite{Damour:2012yf}.

So far, the tidal corrections to the GW phase have been computed only for \emph{nonspinning} objects, i.e., neglecting
the coupling between the angular momentum of one body and the tidal field produced by its companion. In this paper, we
make an important step forward in the PN modeling of the GW signal from spinning NS binaries, by computing the
leading-order tidal interaction of spinning bodies in a binary to leading order in the spin, and the corresponding
corrections to the GW phase. Although the dimensionless spin of NSs in coalescing binaries is expected to be
small~\cite{PhysRevD.86.084017,2013PhRvD..88b1501K}, neglecting the spin-tidal coupling might introduce systematics in
the parameter estimation, especially when using uniform priors that extend to high values of the
spin~\cite{TheLIGOScientific:2017qsa,Harry:2018hke}. Furthermore, spin-tidal corrections might be important to improve
current tests of the nature of compact objects using the tidal effects in the
inspiral~\cite{Wade:2013hoa,Cardoso:2017cfl,Sennett:2017etc,Maselli:2017cmm,Cardoso:2017cqb,Cardoso:2017njb,Johnson-McDaniel2018}.

In recent years, there has been remarkable progress in studying the tidal deformability of spinning compact objects.
Tidal deformations of slowly-spinning black holes were studied in Refs.~\cite{Poisson:2014gka,Pani:2015hfa}, which
confirmed that the TLNs of a black hole are precisely
zero~\cite{Binnington:2009bb,Damour:2009vw,Damour:2009va,Porto:2016zng} also in the spinning case, at least to quadratic
order in the spin in the axisymmetric case and to linear order in the spin in general.  Furthermore, the coupling
between the tidal fields and the angular momentum introduces new families of TLNs, which were dubbed \emph{rotational}
tidal Love numbers
(RTLNs)~\cite{Pani:2015nua,Landry:2015cva,Landry:2015zfa,Landry:2017piv,Gagnon-Bischoff:2017tnz}. While also the RTLNs
of a black holes are precisely zero, those of a NS depend on the equation of state\footnote{We note that the recent
  analysis in Ref.~\cite{Gagnon-Bischoff:2017tnz} found disagreement with the RTLNs previously computed by some of
  us~\cite{Pani:2015nua}, especially for low-compactness stars.  The source of such disagreement is under investigation
  but is irrelevant for the analysis of this work.}.  Finally, a different choice of assumptions on the dynamics of the fluid within the
star (e.g., whether the fluid is irrotational or static) can provide a more realistic configuration for a stationary
tidally distorted object in a binary system, and affect the magnetic TLNs and the
RLTNs~\cite{Landry:2015cva,Landry:2015snx,Gagnon-Bischoff:2017tnz}.

\subsection{Notation and conventions}\label{sec:notation}
We denote the speed of light in vacuum by $c$ and set the gravitational constant $G=1$ throughout the paper.
Latin indices $i,j,k$, etc.\ run over three-dimensional spatial coordinates and are contracted with the Euclidean flat
metric $\delta^{ij}$. Since there is not distinction between upper and lower spatial indices, we will use only the upper
ones throughout the paper. The
complete antisymmetric Levi-Civita symbol is denoted by $\epsilon^{ijk}$. Following the STF notation~\cite{Thorne:1980ru},
we use capital letters in the middle of the alphabet $L,K$, etc.\ as shorthand for multi-indices $a_1 \dots a_l$, $b_1
\dots b_k$, etc. Round $(\ )$, square $[\ ]$, and angular $\langle \ \rangle$ brackets in the indices indicate
symmetrization, antisymmetrization and trace-free symmetrization, respectively. 
For instance,
\begin{equation}
T^{\langle ab \rangle} = T^{(ab)} - \frac{1}{3} \delta^{ab} T^{cc} = \frac{1}{2} \left( T^{ab} + T^{ba} \right) -
\frac{1}{3} \delta^{ab} T^{cc}\,.
\end{equation}
We call \emph{symmetric trace-free} (STF) those tensors $T^{i_1\dots i_l}$ that are symmetric on all indices and
whose contraction of any two indices vanishes
\begin{align}
T^{(i_1\dots i_l)}  &= T^{i_1\dots i_l}\,,\nonumber \\
T^{i_1\dots i_k i_k \dots i_l} &= 0\,, \nonumber\\
T^{\langle i_1\dots i_l \rangle }  &= T^{i_1\dots i_l} \,. 
\end{align}
The contraction of a STF tensor $T^L$ with a generic tensor $U^L$ is $T^L U^L= T^L U^{\langle L \rangle}$. For a generic
vector $u^i$ we define $u^{ij\dots k} \equiv u^i u^j \dots u^k$ and $u^2 \equiv u^{ii}$. Derivatives with respect to the
coordinate time $t$ are expressed by overdots.

For a generic body $A$, the mass and current multipole moments are denoted by $M^L_A$ and $J^L_A$, respectively. We
indicate the electric and magnetic tidal moments, which affect the body $A$, respectively, by $G^L_A$ and $H^L_A$. All
of them are STF tensors on all indices.

Restricted to a two-body system, $A =1,2$, we define the mass ratios $\eta_A = {}^n M_A/M$, where $M = {}^n M_1 + {}^n
M_2 $ is the total mass and ${}^n M_A$ is the mass monopole $M_A$ in the Newtonian limit. The symmetric mass ratio is
$\nu = \eta_1 \eta_2$ and the reduced mass is $\mu = \nu M$. We define the dimensionless spin parameters $\chi_A = c
J_A/(\eta_A M)^2$, where $J_A = \sqrt{J^i_A J^i_A}$ is the absolute value of the current dipole moment.  The body
position, velocity and acceleration vectors are denoted by $z_A^i$, $v_A^i = \dot{z}_A^i$ and $a_A^i = \ddot{z}_A^i$,
respectively. We define the two-body relative position, velocity and acceleration vectors by $z^i = z_2^i - z_1^i$, $v^i
= v_2^i - v_1^i$ and $a^i = a_2^i - a_1^i$, respectively. We also define the relative unit vector $n^i = z^i/r$, where
$r = \sqrt{z^i z^i}$ is the orbital distance. We define the derivatives with respect to the spatial coordinates $z^i$ as
$\partial_L = \partial_{i_1}\dots\partial_{i_l}$. In particular, we denote the derivatives with respect to $z_A^i$ by
$\partial_L^{(A)}$.  We shall also make use of the following identity
\begin{equation}
  \partial_L\frac{1}{r} =\partial_L^{(2)}\frac{1}{r} = (-1)^l
  \partial_L^{(1)} \frac{1}{r} =(-1)^l(2l-1)!!\frac{n^{\langle L\rangle}}{r^{l+1}} \,.
\end{equation}

We shall denote $\lambda_l$ ($\sigma_l$) the electric (magnetic) TLN of multipolar order $l$, whereas $\lambda_{ll'}$
and $\sigma_{ll'}$ are the RTLNs.  As discussed below, for our computation it is sufficient to consider that the multipole moments higher
than the dipole are induced only on the second body by the tidal field produced by its companion. For this
reason, to avoid burdening the notation, we define the quadrupolar and octupolar moments as $Q^{ab} \equiv M_2^{ab}$,
$Q^{abc} \equiv M_2^{abc}$, $S^{ab} \equiv J_2^{ab}$ and $S^{abc} \equiv J_2^{abc}$. To our order of approximation, the
moments induced on object $1$ due to the tidal field produced by object $2$ can be included \emph{a posteriori} by
inverting the indices in the final formulas. We do so only when presenting the final GW phase, Eq.~\eqref{PHASE}.

Finally, for a binary system in circular orbit we define the PN expansion parameter
$x = (\omega M)^{2/3}/c^2$, where $\omega$ is the orbital angular velocity. Note that $x=v^2/c^2+O(c^{-4})$.

\subsection{Tidal deformations of rotating stars}
Finite-size effects due to the deformability of compact objects enter the GW phase through the TLNs. Loosely speaking,
the TLNs can be defined as the multipole moments induced on an object by an external tidal field \emph{per unit} of the
external field itself~\cite{PoissonWill}. Within linear perturbation theory, the TLNs do not depend on the source of the
tidal field but only on the internal properties of the central object.

To linear order in the spin, and assuming small and slowly varying external tidal fields, the TLNs relevant for this
paper can be defined through the following relations: 
\begin{equation}
\begin{split}
Q^{ab}  &=  \lambda_{2} G^{ab}  + \frac{\lambda_{23}}{c^2}   J^c H^{abc}\\
Q^{abc} &=   \lambda_{3} G^{abc} + \frac{\lambda_{32}}{c^2}  J^{\langle c} H^{ab \rangle} \\
S^{ab}  &=   \frac{\sigma_{2}}{c^2} H^{ab}   + {\sigma_{23}} J^c G^{abc}  \\
S^{abc} &=   \frac{\sigma_{3}}{c^2} H^{abc}  + {\sigma_{32}} J^{\langle c} G^{ab \rangle}\,.
\end{split}
\label{eq:adiabatic}
\end{equation} 
These are called {\it adiabatic relations} because the TLNs are assumed to be constant, neglecting the oscillatory
response of the star to a variation of the tidal field (however, see~\cite{Maselli:2012zq,Steinhoff:2016rfi}).

In the above equations, $Q^{L}$ and $S^{L}$ are, respectively, the mass and current multipole moments of order $l$ induced
on the spinning object\footnote{We remind that the index $L$ represents $l$ indices $i_1,\dots,i_l$ running from one to
  three (see Sec.~\ref{sec:notation}).} (with spin vector $J^c$), whereas $G^{L}$ and $H^{L}$ are the external electric
and magnetic tidal moments of order $l$ evaluated at the location of the object. The constants $\lambda_{l}$ and
$\sigma_{l}$ are the ordinary electric and magnetic TLNs, whereas $\lambda_{ll'}$ and $\sigma_{ll'}$ are the
RTLNs~\cite{Pani:2015nua,Landry:2015cva,Landry:2015zfa}. The powers of $c$ in the above equations guarantee that, at
Newtonian order, a magnetic tidal field does not induce any multipole moment. The magnetic tidal
moments source the mass multipole moments only starting at $1$PN order, in agreement with the discussion in
Ref.~\cite{Gagnon-Bischoff:2017tnz}. On the other hand, an electric tidal field can induce also current multipole moments at Newtonian order, but these moments affect the metric only at higher PN order, as discussed below.

The above relations generalize to spinning objects the standard proportionality relations among the quadrupole moments
$(Q^{ab},S^{ab})$ of a nonspinning object and the external quadrupolar tidal moments
$(G^{ab},H^{ab})$~\cite{PoissonWill,Flanagan:2007ix,Hinderer:2007mb}.
In particular, the structure of Eq.~\eqref{eq:adiabatic} corresponds to the spin-tidal couplings introduced in
   Ref.~\cite{Pani:2015nua}:
\begin{itemize}
 \item[(i)] in the nonspinning case, Eq.~\eqref{eq:adiabatic} reduces to $Q^L=\lambda_l G^L$ and $S^L=\frac{\sigma_l}{c^2} H^L$.
   In other words, an $l$-pole tidal moment can induce only an $l$-pole multipole moment
   with the same parity\footnote{Electric and magnetic tidal moments have even and odd parity,
     respectively. Likewise, mass and current multipole moments have even and odd parity, respectively.}. For example, a
   quadrupolar electric (respectively, magnetic) tidal moment $G^{ab}$ (respectively, $H^{ab}$) induces a mass (respectively, current)
   quadrupole moment, $Q^{ab}$ (respectively, $S^{ab}$). For $l=2$ and $l=3$, the induced multipole moments depend on four
   independent TLNs, namely $\lambda_{2}$, $\lambda_{3}$, $\sigma_{2}$, and $\sigma_{3}$. It is well known
   that the dominant correction to the GW phase depends on $\lambda_{2}$ through a $5$PN
   term~\cite{Flanagan:2007ix,Hinderer:2007mb};
 \item[(ii)] the spin of the binary components couples tidal moments and multipole moments with different $l$ order and
   opposite parity~\cite{Pani:2015nua,Pani:2013pma}. In particular, a \emph{magnetic} quadrupolar (respectively, octupolar)
   tidal moment can induce a \emph{mass} octupole (respectively, quadrupole) moment through a term proportional to the spin and
   to the rotational Love number $\lambda_{32}$ (respectively, $\lambda_{23}$). Likewise, an \emph{electric} quadrupolar
   (respectively, octupolar) tidal moment can induce a \emph{current} octupole (respectively, quadrupole) moment through a term
   proportional to the spin and to the rotational Love number $\sigma_{32}$ (respectively, $\sigma_{23}$).
\end{itemize}

\subsection{Summary of results}\label{subsec:summrel}
For the busy reader, we summarize here the main results of our work, which are derived in detail in the rest of the
paper. We follow the notation described in Sec.~\ref{sec:notation}.  Our main result is the GW phase with \emph{all}
tidal corrections included up to $6.5$PN order and to linear order in the spin, see Eq.~\eqref{PHASE} below.

\subsubsection{Lagrangian}

The Lagrangian describing the two-body interaction can be written as
\begin{equation}
\label{eq:totallagrangian}
\mathcal{L} = \mathcal{L}_{orb} + \mathcal{L}_2^{int}\,. 
\end{equation}
Here, $\mathcal{L}_{orb}$ describes the orbital motion of the bodies 
\begin{equation}
  \mathcal{L}_{orb} = \mathcal{L}_M + \mathcal{L}_J + \mathcal{L}_{Q2} +
  \mathcal{L}_{Q3} + \mathcal{L}_{S2}  + \mathcal{L}_{S3}\,, \label{Lorb}
\end{equation}
where $\mathcal{L}_M$ and $\mathcal{L}_J$ are the contributions that depend only on the masses of the
two bodies and on their spin vectors (to linear order in the spin), respectively,
\begin{align}
\mathcal{L}_M =& \frac{\mu v^2}{2} + \frac{\mu M}{r} + \frac{\mu}{c^2}
\left\{\frac{1-3 \nu}{8}v^4\right.\nonumber\\
  &\left.+ \frac{M}{2r} \left[\left(3+ \nu\right)v^2+ \nu
  \dot{r}^2-\frac{M}{r} \right] \right\} +O \left( c^{-4} \right)\,,\label{eq:LM}\\
\mathcal{L}_J =& \frac{\epsilon^{abc}}{c^2}   v^b \left[ \left( \eta_2 J_1^a
  + \eta_1 J_2^a \right)  \frac{2M}{r^2} n^c  \right. \nonumber\\
&\left. + \left( \eta_2^2 J_1^a + \eta_1^2 J_2^a  \right) \frac{a^c}{2}  \right] +O \left( c^{-4} \right)
  \,.\label{eq:LJ}
\end{align}
The mass quadrupole term reads, to next-to-leading PN order,
\begin{widetext}
\begin{align}
  \mathcal{L}_{Q2} =& \frac{3 \eta_1 M}{2r^3} Q^{ab} n^{ab} + \frac{1}{c^2} \bigg\{ \frac{M}{r^3} Q^{ab} \bigg[ n^{ab}
    \bigg( \frac{3 \eta_1}{4} (3+\nu) v^2 + \frac{15 \nu \eta_1}{4} \dot{r}^2 -\frac{3 \eta_1}{2} (1+3\eta_1)\frac{M}{r} \bigg)  +\frac{3 \eta_1^2}{2} v^{ab}     -\frac{3 \eta_1^2}{2}(3+\eta_2) \dot{r} n^a v^b \bigg] \nonumber\\
&  - \frac{M}{r^2} \dot{Q}^{ab} \left[\frac{3 \nu }{2} n^a v^b +\frac{3 \nu }{4}\dot{r} n^{ab}\right]
  +E_2^{int} \left[\frac{\eta_1^2}{2} v^2 + \eta_1 \frac{M}{r} \right] \bigg\}  + \frac{3 }{c^2 r^4} \epsilon^{icd} J_1^a Q^{bd} \left(5 n^{abc}
  -\delta^{ab} n^{c} -\delta^{ac} n^b \right) v^i +O \left( c^{-4} \right) \,,
  \label{eq:LMQ}
\end{align}
\end{widetext}
where $E_2^{int}$ is the internal energy of body $2$ which, to this level of approximation, can be expressed through its
Newtonian value given in Eq.~\eqref{EintN} below.
At Newtonian order, $\mathcal{L}_{Q2}$ is simply a coupling between the mass quadrupole moment and the quadrupolar
electric tidal moment, $\mathcal{L}_{Q2}=\frac{1}{2}G_2^{ab}Q^{ab}+O(c^{-2})$.

Likewise, the mass octupole term reads, to leading order,
\begin{equation}
\label{eq:LQ3}
\mathcal{L}_{Q3}  = \frac{1}{6}G_2^{abc}Q^{abc}\equiv -\frac{5 \eta_1 M}{2r^4} Q^{abc} n^{abc} +O \left( c^{-2} \right)\,,
\end{equation}
whereas the current quadrupole and octupole terms, respectively, read (again to leading order)
\begin{align}
\mathcal{L}_{S2}  =& \frac{1}{3c^2}H_2^{ab}S^{ab}\nonumber\\
\equiv &  \frac{4 \eta_1 M}{c^2 r^3} \epsilon^{bcd} n^{ab} S^{ad} v^c + \frac{2}{c^2 r^4} J_1^c S^{ab} 
\left( 5 n^{abc} -2 \delta^{bc} n^a  \right)\nonumber\\
&+O \left( c^{-4} \right) \,, \label{eq:LS2}
\end{align}
and
\begin{align}
\mathcal{L}_{S3}  =& \frac{1}{8c^2}H_2^{abc}S^{abc} \nonumber \\
\equiv &  \frac{15 \eta_1 M}{2c^2r^4} \epsilon^{ade} S^{bcd}  n^{abc}v^e 
 \nonumber \\
&+\frac{45}{4 c^2 r^5} J_1^d S^{abc} \left(\delta^{cd} n^{ab} -
\frac{7}{3} n^{abcd} \right) +O \left( c^{-4} \right)\,.
\label{eq:LS3}
\end{align}
The other term appearing in Eq.~\eqref{eq:totallagrangian}, $\mathcal{L}_2^{int}$, takes into account the internal
dynamics of the body 2 (which we remind is the only one tidally deformed at this stage),
\begin{align}
  \mathcal{L}_2^{int} =& -\frac{1}{4 \lambda_{2}} Q^{ab} Q^{ab} -\frac{1}{12 \lambda_{3}}
  Q^{abc} Q^{abc} -\frac{1}{6 \sigma_{2}} S^{ab} S^{ab}\nonumber \\
  &-\frac{1}{16 \sigma_{3}} S^{abc} S^{abc} +\alpha J_2^a Q^{bc} S^{abc} +
  \beta J_2^a S^{bc} Q^{abc}\,, \label{L2int}
\end{align}
where ${\alpha}$ and ${\beta}$ are related to the RTLNs in a way to be defined shortly.  Those above are all possible
couplings among the multipole moments included in our model, and by the requirement that $\mathcal{L}_2^{int}$ is
scalar, parity invariant, at most linear in the spin, and quadratic in the higher multipole moments.

The Lagrangian~\eqref{Lorb} provides the correct equations of motion to linear order in the spin, up to $1$PN order in
the electric quadrupolar TLN, and to leading order in the other tidal deformations. As discussed below, this is
sufficient to completely describe the tidal contribution to the phase up to $6.5$PN order.
Furthermore, the interaction term~\eqref{L2int} guarantees that Euler-Lagrange equations for the multipole moments
yield the adiabatic relations~\eqref{eq:adiabatic} with the following identification:
 \begin{align}
   \lambda_{23}&= 2 \lambda_{2} \sigma_{3} \alpha \,\,\,\qquad
   \lambda_{32}= 6  \lambda_{3} \sigma_{2} \beta  \nonumber \\
   \sigma_{23}&= 3  \lambda_{3} \sigma_{2}\beta \,\,\, \qquad
   \sigma_{32}= 8   \lambda_{2} \sigma_{3}\alpha  \,. \label{rotTLNs}
 \end{align}
We note that, with the above definition, the RTLNs are proportional to those defined in Ref.~\cite{Pani:2015nua} in the
axisymmetric case. The explicit relations between them are given in Appendix~\ref{app:comparison}.

Crucially, the Lagrangian formulation enforces Eq.~\eqref{rotTLNs}. Therefore, only \emph{two out of four} RTLNs are
independent. In particular, $\lambda_{23}$ is proportional to $\sigma_{32}$ and $\lambda_{32}$ is proportional to
$\sigma_{23}$. This proportionality does not emerge from the perturbative computation performed in
Ref.~\cite{Pani:2015nua}. We discuss this issue in more detail in Sec.~\ref{sec:issue}.

\subsubsection{GW phase}

Finally, from the above Lagrangian one can compute the GW phase to linear order in the spin. For circular orbits, up
to $1.5$PN order in the point-particle terms and up to $6.5$PN order in the tidal-deformability terms, the GW phase 
for the TaylorF2 approximant~\cite{Arun:2008kb,Buonanno:2009zt,Mishra:2016whh} reads
\begin{align}
&  \psi(x)  =  \frac{3}{128 \nu x^{5/2}} \Bigg\{ 1 + \left(\frac{3715}{756}+\frac{55}{9} \nu \right) x+
  \left(\frac{113}{3}\right.\times\nonumber\\
  &\times\left.(\eta_1 \chi_1 + \eta_2 \chi_2)  -\frac{38}{3} \nu (\chi_1 + \chi_2 ) -16 \pi \right) x^{1.5} +O(x^2) \nonumber\\
  & + \Lambda  x^5 + (\delta\Lambda+\Sigma) x^6 + (\tilde\Lambda+\tilde\Sigma+ \tilde\Gamma-\pi \Lambda) x^{6.5} +{ O}(x^7)\Bigg\}\,,
  \nonumber\\
\label{PHASE}
\end{align} 
where $x=\frac{1}{c^2}(M\omega)^{2/3}$, $\omega$ is the orbital angular velocity, and $\chi_i$ are the dimensionless spin
parameters introduced in Sec.~\ref{sec:notation}. 

The first two lines in the above equation denote the point-particle contribution to the GW phase, whereas the other
terms are due to the tidal deformability. The $5$PN term is the usual leading-order tidal contribution where
\begin{equation}
 \Lambda =\left(264 -\frac{288}{\eta_1}\right) \frac{c^{10}\lambda_2^{(1)}}{M^5} +
    (1\leftrightarrow 2)\,,
\end{equation}
and $\lambda_2^{(A)}$ is the (quadrupolar, electric) TLN of the $A$-th body.  The $6$PN term contains the
next-to-leading order contribution of the previous term,
\begin{align}
 \delta\Lambda=&\left(  \frac{4595}{28}- \frac{15895}{28 \eta_1}  + \frac{5715 \eta_1}{14} -
 \frac{325 \eta_1^2}{7}   \right)\frac{c^{10}\lambda_2^{(1)}}{M^5}\nonumber\\
 &+(1\leftrightarrow 2) \,,
\end{align}
and the leading-order contribution from the quadrupolar, \emph{magnetic} TLN,
\begin{equation}
 \Sigma =  
    \left( \frac{6920}{7} - \frac{20740}{21 \eta_1} \right) \frac{c^{8}\sigma_2^{(1)}}{M^5}+(1\leftrightarrow 2)\,.
\end{equation}
Note that the magnetic term corrects some errors in the first version of Ref.~\cite{Yagi:2013sva} and agrees with that
recently derived in Ref.~\cite{Banihashemi:2018xfb}. In particular, $\Sigma$ affects the phase at $6$PN order also for
equal-mass binaries and is therefore degenerate with $\delta\Lambda$.\footnote{The contribution from the quadrupolar
  magnetic TLN is typically ignored since it was thought to enter at higher PN order for circular binaries. At any rate,
  even if this term is degenerate with $\delta\Lambda$, $\sigma_2$ is typically smaller than
  $\lambda_2$~\cite{Binnington:2009bb,Landry:2015cva}. The contribution from $\Sigma$ is small but potentially
  detectable~\cite{Jimenez-Forteza:2018buh}.} The leading-order tail-tidal term, proportional to the quadrupolar electric TLN
$\lambda_2$, enters in the GW phase at $6.5$PN order with the same combination $\Lambda$ as in the $5$PN
term~\cite{Damour:2012yf}.

Finally, all spin-tidal terms enter at $6.5$PN order in the GW phase, through three different (albeit degenerate)
terms. The first two are due to the coupling between the spin and the ordinary quadrupolar (electric and magnetic)
TLNs:
\begin{align}
  \tilde\Lambda =&
  \left[ \left( \frac{593}{4} - \frac{1105}{8 \eta_1}
    +\frac{567 \eta_1}{8} -81 \eta_1^2 \right) \chi_2\right. \nonumber\\
     &+\left.\left( -\frac{6607}{8} +\frac{6639 \eta_1}{8} -81 \eta_1^2 \right)
     \chi_1 \right] \frac{c^{10}\lambda_2^{(1)}}{M^5}\nonumber \\
   &+(1\leftrightarrow 2)\,, \label{tildeLambdaE}\\
  \tilde\Sigma =&\left[\left(-\frac{9865}{3} + \frac{4933}{3 \eta_1} + 1644 \eta_1 \right) \chi_2 -\chi_1 \right]
  \frac{c^{8}\sigma_2^{(1)}}{M^5}\nonumber \\
  &+(1\leftrightarrow 2)\,, \label{tildeLambdaM}
\end{align}
whereas the third $6.5$PN term in Eq.~\eqref{PHASE} is proportional to the RTLNs,
\begin{align}
  \tilde\Gamma =&\frac{c^{10} \chi_1}{M^4}\left[ \left(   856 \eta_1
    -  816 \eta_1^2 \right){\lambda_{23}^{(1)}} \right.\nonumber \\
  &\left. - \left(\frac{833 \eta_1}{3} - 278 \eta_1^2  \right) {\sigma_{23}^{(1)}}\right. \nonumber\\
& \left. - \nu \left(272 {\lambda_{32}^{(1)}} -204 {\sigma_{32}^{(1)} }\right)\right]+(1\leftrightarrow 2)\,.
  \end{align}

The fact that these terms all enter at $6.5$PN order can be understood as follows. Let us first focus on the terms
proportional to the RTLNs, i.e., on $\tilde\Gamma$.  The mass quadrupole moment $Q^{ab}$ enters at $2$PN order in the
phase~\cite{Poisson:1997ha,Mikoczi:2005dn,Blanchet:2006zz}. From the adiabatic relations~\eqref{eq:adiabatic}, $Q^{ab}$
acquires a contribution proportional to the spin and to $H^{abc}\sim v/r^4\sim 4.5{\rm PN}$, so that overall these
corrections enter the GW phase at $2+4.5=6.5$PN order. Likewise, $Q^{abc}$ affects the phase at $3$PN order and its
spin-tidal coupling is proportional to $H^{ab}\sim v/r^3\sim 3.5{\rm PN}$ so that also this term enters at $3+3.5=6.5$PN
order. Similar arguments can be made for the terms proportional to $S^{ab}$ and $S^{abc}$, since the latter enter the GW
phase at $2.5$PN and $3.5$PN order, respectively, and they are coupled to $G^{abc}\sim 4$PN and $G^{ab}\sim3$PN,
respectively.

We generalize this argument in Sec.~\ref{subsec:counting}, showing that the spin-tidal terms arising from $l$-pole RTLNs
to linear order in the spin enter the GW phase at $(2l+1/2+2\delta_{l2})$PN order. Therefore, for any $l\geq3$, this
contribution enters at \emph{lower} PN order relative to the standard electric TLNs of order $l$ (the latter entering at
$(2l+1)$PN order).

Corrections proportional to the spin and to the ordinary $l=2$ TLNs, i.e. the $\tilde\Lambda$ and $\tilde\Sigma$ terms,
also enter at $6.5$PN order. This is due to the fact that, as discussed below, the leading-order spin terms in $G^{ab}$
and $H^{ab}$ enter, respectively, at $4.5$PN and at $4$PN order and, since they enter the GW phase, respectively, through
the induced $Q^{ab}$ and $S^{ab}$ (at $2$PN and $2.5$PN order, respectively), their overall contribution is again $6.5$PN.
We generalize this argument in Sec.~\ref{subsec:counting}, showing that the spin-tidal terms arising from $l$-pole TLNs
(both electric and magnetic) to linear order in the spin enter the GW phase at $(2l+5/2)$PN order.  It is also worth
noting that these terms effectively couple higher-order point-particle terms (the spins) to the tidal terms (the
ordinary TLNs), thus breaking the ``decoupling'' that exists between the point-particle phase and the tidal phase at the
leading order~\cite{Flanagan:1997fn}.

The tidal terms entering the GW phase at leading order in the spin are summarized in Table~\ref{tab:tidal}. For
completeness, in Appendix~\ref{app:higher} we also provide higher-order terms entering the GW phase~\eqref{PHASE} which
are proportional to the TLNs and are computed as a by-product of our analysis.

We stress that --~within our Lagrangian approach~-- only two out of $\lambda_{32}$, $\lambda_{23}$, $\sigma_{32}$, and
$\sigma_{23}$ are independent, these four quantities being related to ${\alpha}$ and ${\beta}$ [the only two extra
  parameters entering our interaction Lagrangian~\eqref{L2int}] through Eq.~\eqref{rotTLNs}. While this is an unsolved
issue (see discussion in Sec.~\ref{sec:issue}) for the sake of generality we will consider these terms as
independent. In any case, these terms enter the GW phase only through the combination~$\tilde\Gamma$.

\begin{table}
 \caption{Schematic representation of the PN contributions of the TLNs and of the RTLNs to the GW phase of a binary
   system to linear order in the spin. ``LO'', ``NLO'', and ``NNLO'' stand for Leading Order, Next-to-Leading Order,
   etc. The entries in boldface are the new $6.5$PN terms computed in this work (we omit the leading-order tail effect entering at $6.5$PN order derived in~\cite{Damour:2012yf}). They are all proportional to the spins
   of the binary components and would be zero in the nonspinning case.  For generic $l$-poles, the contribution from
   RTLNs enters at $(2l+1/2+2\delta_{l2})$PN order, whereas the spin-tidal contribution from the ordinary TLNs enters at
   $(2l+5/2)$PN order (see Sec.~\ref{subsec:counting}). For comparison, in the nonspinning case the electric and
   magnetic TLNs enter at $(2l+1)$PN and $(2l+2)$PN order, respectively.}
 \begin{tabular}{|c||c|c|c|c|c|}
 \hline
 \hline
 PN order & $\lambda_{2}$ & $\sigma_{2}$ & $\lambda_{23,32}$,  $\sigma_{23,32}$  & $\lambda_{3}$ & $\sigma_{3}$ \\ 
 \hline
 $5$ 	& LO $\propto\Lambda$   &      &     	  &     &     \\
 $6$ 	& NLO $\propto\delta\Lambda$  & LO $\propto\Sigma$   &     	  &     &     \\
 $\bm{6.5}$  & {\bf NNLO} $\bm{\propto\tilde\Lambda}$     & {\bf NLO} $\bm{\propto\tilde\Sigma}$
 & {\bf LO} $\bm{\propto\tilde\Gamma}$   &    &     \\
 $7$ 	& $\dots$ & $\dots$  &  $\dots$   	  & LO  &     \\
 $8$ 	& $\dots$ & $\dots$  &  $\dots$   	  & $\dots$  &  LO   \\
 \hline
 \hline
 \end{tabular}
   \label{tab:tidal}
\end{table}

\section{PN tidal interactions of  interacting, structured bodies}\label{sec:pntidal}
In this section, we summarize the PN theory of tidal interactions in binary systems, which has been mainly
developed in Refs.~\cite{Damour:1991prd,Racine:2004hs,Vines:2011ud,Vines:2013prd}.
\subsection{Coordinate frames and multipole expansions of the PN potentials}

Let us consider $N$ interacting, arbitrarily structured bodies immersed in a strong-field environment.  It is
possible to define a harmonic\footnote{The harmonic gauge condition is $\partial_{\mu}(\sqrt{-g} g^{\mu \nu})=0$, which
  implies $4 \dot{\Phi}+ \partial_i \zeta^i = O(c^{-2})$.} and conformally Cartesian\footnote{Conformally Cartesian
  coordinates are a special case of isotropic coordinates and require $g_{00} g_{ij} = -\delta_{ij} +
  O(c^{-4})$~\cite{Damour:1991prd}.} coordinate system $(t,x^i)$, which we call ``global frame'', covering the entire
spacetime except the strong-field region near each body. In this frame, the spacetime metric, in $1$PN approximation
[i.e., including terms up to $O(c^{-2})$] has the form
\begin{align}
\label{eq:1pn_metric}
ds^2 &= - \left(1+ \frac{2 \Phi_g}{c^2} + \frac{2 \Phi_g^2}{c^4} \right) c^2 dt^2
+ \frac{2 \zeta_g^i}{c^3} c dt dx^i \nonumber\\
&+ \left(1-\frac{2 \Phi_g}{c^2} \right)
\delta^{ij} dx^i dx^j + O( c^{-4})\,.
\end{align}
The scalar field $\Phi_g(t,\boldsymbol{x})$ can be decomposed as $\Phi_g=\phi_g+c^{-2}\psi_g$, where $\phi_g$ is the
Newtonian ($0$PN) potential and $\psi_g$ is its $1$PN correction; $\zeta_g^i(t,\boldsymbol{x})$ is the
gravito-magnetic vector potential, at $1$PN order.

For each body $A$ ($A=1,\dots,N$) we assume the existence of a local coordinate system $(s_A,y^i_A)$, which we call
``body frame'' or ``local frame'', covering the body, including the strong-field worldtube $\mathcal{W}_A$ defined as
the product of the ball $|\boldsymbol{y}_A|<r_{A\,-}$ with an open interval of time. Moreover, for each body $A$, there
exists a buffer region $\mathcal{B}_A$ defined as the product of $r_{A\,-}<|\boldsymbol{y}_A|<r_{A\,+}$ with an open
time interval, which is covered by both the global frame $(t,x^i)$ and the local frame $(s_A,y^i_A)$. In the buffer
region $\mathcal{B}_A$, the gravitational field is weak and the local coordinates $s_A,y^i_A$ are harmonic and
conformally Cartesian, therefore the metric can be written in the $1$PN form shown in Eq.~\eqref{eq:1pn_metric}, in
terms of potentials $\Phi_A(s_A,\boldsymbol{y}_A)$, $\zeta^i_A(s_A,\boldsymbol{y}_A)$. In the buffer region, the
coordinate transformation between the global frame and the local frame has the form
\begin{equation}
x^i(s_A,\boldsymbol{y}_A)=y^i_A+z^i(s_A)+c^{-2}\left[1\hbox{PN terms}\right]\,,
\label{eq:coordtransf}
\end{equation}
where the vector $z^i$ describes a time-dependent spatial translation between the two frames.  
We do not explicitly write the $1$PN terms in Eq.~\eqref{eq:coordtransf} for brevity; they depend on a set of freely
specifiable functions encoding the residual gauge freedom.

As shown in Ref.~\cite{Racine:2004hs}, under these assumptions the potentials $\Phi_A$, $\zeta^i_A$ for each body can be
written in terms of a set of {\it multipole moments}.  The internal degrees of freedom of the body are described by its
{\it mass multipole moments} $M_A^L(s_A)$ (with $l\ge0$) and its {\it current multipole moments} $J_A^L(s_A)$ (with
$l\ge1$)~\cite{Geroch:1970cd,Hansen:1974zz,Thorne:1980ru} (for a recent account in the context of tests of the
black-hole no-hair theorem, see also~\cite{Cardoso:2016ryw}).  The tidal field due to the bodies $B\neq A$ is described
by the electric tidal moments $G_A^L(s_A)$ and the magnetic tidal moments $H_A^L(s_A)$ (defined for
$l\ge0$ and $l\ge1$, respectively). Both the body and the tidal  moments are STF tensors on all indices. The
explicit expansion of the PN potentials in the body frame is
\begin{align}
  \Phi_A(s_A,\boldsymbol{y}_A)=& -\sum_{l=0}^\infty\frac{1}{l!}\left\{(-1)^lM_A^L(s_A)
  \partial_L\frac{1}{|\boldsymbol{y}_A|}\right.\nonumber\\
  &+G_A^L(s_A)y_A^L\nonumber\\
&+\frac{1}{c^2}\left[\frac{(-1)^l(2l+1)}{(l+1)(2l+3)}{\dot\mu}^L_A(s_A)
  \partial_L\frac{1}{|\boldsymbol{y}_A|}\right.\nonumber\\
  &+\frac{(-1)^l}{2}{\ddot M}^L_A(s_A)\partial_L|\boldsymbol{y}_A|-{\dot\nu}^L_A(s_A)y^L_A\nonumber\\
  &\left.\left.+\frac{1}{2(2l+3)}{\ddot G}^L_A(s_A)y_A^{jjL}\right]\right\}+O(c^{-4}) \,, \nonumber\\
\zeta^i_A(s_A,\boldsymbol{y}_A)=& -\sum_{l=0}^\infty\frac{1}{l!}\left\{(-1)^lZ_A^{iL}(s_A)
\partial_L\frac{1}{|\boldsymbol{y}_A|}\right. \nonumber \\
&+Y_A^{iL}(s_A)y_A^L\bigg\}+O(c^{-2})\,,\label{eq:pn_expansion}
\end{align}
where
\begin{align}
Z^{iL}_A(s_A)=&\frac{4}{l+1}{\dot M}^{iL}_A(s_A)-\frac{4l}{l+1}\epsilon^{ji \langle a_l}J_A^{L-1 \rangle j}(s_A)\nonumber\\
&+\frac{2l-1}{2l+1}\delta^{i\langle a_l}\mu_A^{L-1 \rangle}(s_A)+O(c^{-2})\,,\\
Y_A^{iL}(s_A)=&\nu_A^{iL}(s_A)+\frac{l}{l+1}\epsilon^{ji \langle a_l}H_A^{L-1 \rangle j}(s_A)\nonumber\\
&-\frac{4(2l-1)}{2l+1}{\dot G}_A^{\langle L-1}\delta^{a_l \rangle i}(s_A)+O(c^{-2})\,.
\end{align}
Mass and electric moments are defined up to $1$PN order, while current and magnetic moments are defined just to
Newtonian level.  The quantities $\mu^L_A$, $\nu^L_A$ (defined for $l \geq 0$ and $l\geq 1$, respectively, and not to be
confused with symmetric mass ratio $\nu$ and reduced mass $\mu$) are called internal and external gauge moments,
respectively, because they do not contain gauge-invariant information. As we shall see below, they will be set to zero
by choosing the body-frame coordinate system.

In Eq.~\eqref{eq:pn_expansion}, the separation between the interior and exterior degrees of freedom is clear and unique:
the terms with negative powers of $|\boldsymbol{y}_A|$ depend on the body multipole moments; the terms with positive
powers of $|\boldsymbol{y}_A|$ depend on the tidal  moments. This expression is defined in the buffer region
where $r_{A\,-}<|\boldsymbol{y}_A|<r_{A\,+}$; the body multipole moments encode the structure of the strong-field region
$|\boldsymbol{y}_A|<r_{A\,-}$, while the tidal moments encode the gravitational fields generated by external
($|\boldsymbol{y}_A|>r_{A\,+}$) sources and the inertial effects due to the motion of the local asymptotic rest frame
with respect to the global frame\footnote{In general, discriminating between interior and exterior degrees of freedom
  may be more subtle (see, e.g., Refs.~\cite{Pani:2015hfa,Gralla:2017djj}).}.

Using the residual gauge freedom in the coordinate transformation~\eqref{eq:coordtransf}, we choose the {\it
  body-adapted gauge} for the local frame, by setting $M^i_A=0$ and $G_A=0$; the former ensures that the center of
mass-energy of body $A$ is at $y^i_A=0$, the latter that replacing the body by a freely falling observer at $y^i_A=0$,
the proper time is measured by the coordinate $s_A$. Moreover, we set to zero the internal and external gauge moments,
$\mu_A^L=\nu_A^L=0$, and we choose the orientation of the body-frame spatial axes to coincide with those of the global
frame (see~\cite{Racine:2004hs,Vines:2013prd}). 

In the body-adapted gauge, the coordinate transformation~\eqref{eq:coordtransf} yields a function $z^i_A(t)$ such that
the equation $x^i=z^i_A(t)$ describes the position of the body $A$ in the global frame. This is called the
``center-of-mass worldline'' of the body $A$, but in general it does not parametrize an actual worldline in spacetime
(the global frame $(t,x^i)$ is not defined in the strong-field region of the body, and thus it is not defined in its center of
mass). This function parametrizes the location of the local frame of the body $A$ in the global coordinate system. The
same procedure gives the functions $s_A(t)$ relating the proper and coordinate times of each body.

The PN potentials of the global frame~\eqref{eq:1pn_metric} can be expressed in terms of the {\it global multipole
  moments} of the different bodies $M^L_{g,A}(t)$, $Z^{iL}_{g,A}(t)$ (defined for $l \geq 0$ and STF tensors on all (the
last) $l$-indices):
\begin{align}
  \Phi_g(t,\boldsymbol{x})=&-\sum_{A=1}^{N}\sum_{l=0}^\infty\frac{(-1)^l}{l!}\left\{
    M_{g,A}^L(t)\partial_L\frac{1}{|\boldsymbol{x}-\boldsymbol{z}_A(t)|}\right.\nonumber\\
  &\left.+\frac{1}{2c^2}\partial^2_t\left[M_{g,A}^L(t)\partial_L|\boldsymbol{x}-\boldsymbol{z}_A(t)|\right]\right\}
\nonumber\\
&    + O\left(c^{-4}\right)\,,\nonumber\\
  \zeta_g^i(t,\boldsymbol{x})=&-\sum_{A=1}^{N}\sum_{l=0}^\infty\frac{(-1)^l}{l!}Z^{iL}_{g,A}(t)
  \partial_L\frac{1}{|\boldsymbol{x}-\boldsymbol{z}_A(t)|}\nonumber\\
  &+ O\left(c^{-2}\right)\,.\label{eq:globmultexp}
\end{align}
This expansion is the sum of contributions from each body, but it is defined in the entire spacetime (excluding the
strong-field regions only). The contribution from the body $A$ is centered at $x^i=z^i_A(t)$, and is parametrized by the
global multipole moments $M^L_{g,A}$, $Z^{iL}_{g,A}$. There are no tidal terms, since there is no ``exterior'' tidal
contribution.

In the buffer region of body $A$, the PN potentials $\Phi_g(t,\boldsymbol{x})$, $\zeta^i_g(t,\boldsymbol{x})$ can be
expanded in a different way, including the contributions from the bodies $B\neq A$ in the (electric and magnetic)
global-frame tidal moments $G^L_{g,A}(t)$, $Y^{iL}_{g,A}(t)$ (defined for $l \geq 0$ and STF tensors on all (the
last) $l$-indices):
\begin{align}
  \Phi_g(t,\boldsymbol{x})=&-\sum_{l=0}^\infty\frac{1}{l!}\left\{
   (-1)^l  M_{g,A}^L(t)\partial_L\frac{1}{|\boldsymbol{x}-\boldsymbol{z}_A(t)|}\right.\nonumber\\
    &\left.+G^L_{g,A}(t)[x-z_A(t)]^L  \right.\nonumber\\
    & \left.+\frac{1}{2c^2}\partial^2_t
    \bigg[M_{g,A}^L(t)\partial_L|\boldsymbol{x}-\boldsymbol{z}_A(t)|\right.\nonumber\\
    &\left.\left.+\frac{1}{2l+3}G_{g,A}^L(t)[x-z_A(t)]^{jjL}\right]\right\}\nonumber\\
&  + O\left(c^{-4}\right)\,,\nonumber\\
  \zeta_g^i(t,\boldsymbol{x})=&-\sum_{l=0}^\infty\frac{1}{l!}\left\{(-1)^lZ^{iL}_{g,A}(t)
  \partial_L\frac{1}{|\boldsymbol{x}-\boldsymbol{z}_A(t)|}\right.\nonumber\\
  &+Y^{iL}_{g,A}(t)[x-z_A(t)]^L\bigg\}+ O\left(c^{-2}\right)\,.\label{eq:globlocmultexp}
\end{align}
The global multipole moments $M^L_{g,A}$, $Z^{iL}_{g,A}$ can be written in terms of the body-frame multipole moments of
mass and current, $M^L_A(s_A)$ and $J^L_A(s_A)$. The explicit expressions are given in Appendix~\ref{app:momtransf},
together with the expressions of the body-frame tidal moments $G_A^L$, $H_A^L$ in terms of the global tidal moments.

Far away from all sources, the PN potentials of the global frame can also be expanded in terms of the multipole moments
of the entire source, the {\it system multipole moments} of mass $M^L_{sys}(t)$ and current $J^L_{sys}(t)$ (STF tensors
defined for $l \geq 0$ and $l \geq 1$, respectively):
\begin{align}
  \Phi_g(t,\boldsymbol{x})=&-\sum_{l=0}^\infty\frac{(-1)^l}{l!}\left\{
    M_{sys}^L(t)\partial_L\frac{1}{|\boldsymbol{x}|}\right.\nonumber\\
    &\left.+\frac{1}{c^2}\left[\frac{(2l+1)}{(l+1)(2l+3)}{\dot\mu}^L_{sys}
      \partial_L\frac{1}{|\boldsymbol{x}|}\right.\right.\nonumber\\
      &\left.\left.+\frac{1}{2}{\ddot M}^L_{sys}\partial_L|\boldsymbol{x}|\right]\right\}\nonumber\\
   &+  O\left(c^{-4}\right)\,,\nonumber\\
    \zeta_g^i(t,\boldsymbol{x})=&-\sum_{l=0}^\infty\frac{(-1)^l}{l!}
    Z_{sys}^{iL}(t)\partial_L\frac{1}{|\boldsymbol{x}|}+  O\left(c^{-2}\right)\,\label{eq:globmultexp2}\,,
\end{align}
where $\mu^L_{sys}=Z^{jjL}_{sys}$ and
\begin{equation}
J^L_{sys}=\frac{1}{4}Z_{sys}^{jk \langle L-1}\epsilon^{a_l \rangle kj}\,. \label{sysS}
\end{equation}
They are related to the global multipole moments of the bodies by the relations~\cite{Vines:2013prd,Vines:2011ud}
\begin{align}
  M^L_{sys} =&  \sum_{A=1}^{N} \sum_{k = 0}^l  \binom{l}{k} \bigg[ M_{g,A}^{\langle L-K} z_A^{K \rangle}
   \nonumber\\
   & + \frac{1}{c^2} \frac{1}{2(2l+3)}\partial_t^2 \left(2 M_{g,A} ^{j \langle L- K } z_A^{K \rangle j}\right.\nonumber\\
  & \left.\left.+ M_{g,A}^{\langle L -K} z_A^{K \rangle jj} \right) \right]\nonumber \\
& - \frac{1}{c^2} \frac{2l +1}{(l+1)(2l+3)} \dot{\mu}^L_{sys} +  O\left(c^{-4}\right)\,,\label{sysM}\\
  Z^{iL}_{sys} =&  \sum_{A=1}^{N} \sum_{k=0}^l  \binom{l}{k}
  Z_{g,A}^{i \langle L -K} z_A^{K \rangle}+  O\left(c^{-2}\right) \,.\label{sysZ}
\end{align}

\subsection{Equations of motion}\label{subsec:eom2}
The equations of motion of the body $A$ are the orbital equation of motion [i.e., a differential equation for
$z^i_A(t)$], and the multipole equations of motion (i.e., a set of differential equations for the lowest-order body-frame
multipole moments).

The multipole equations of motion have been derived in Refs.~\cite{Damour:1991prd,Racine:2004hs}, and have the form:
\begin{align}
  \dot{M}_A  =&  -\frac{1}{c^2} \sum_{l=0}^{\infty} \frac{1}{l!} \left[(l+1) M^L_A \dot{G}^L_A
    + l \, \dot{M}^L_A G^L_A \right] \nonumber\\
  &+ O\left(c^{-4}\right) \,, \label{eq:monopole}\\
  \ddot{M}^i_A  = &\sum_{l=0}^{\infty} \frac{1}{l!} \bigg\{ M^L_A G^{iL}_A+ \frac{1}{c^2}
  \bigg[ \frac{1}{l+2} \epsilon^{ijk} M_A^{jL} \dot{H}^{kL}_A \nonumber\\
    &+ \frac{1}{l+1} \epsilon^{ijk} \dot{M}_A^{jL} H^{kL}_A \nonumber\\
    &- \frac{2l^3+7l^2+15l+6}{(l+1)(2l+3)} M^{iL}_A \ddot{G}^L_A\nonumber \\
    & - \frac{2l^3+5l^2+12l+5}{(l+1)^2} \dot{M}^{iL}_A \dot{G}^L_A\nonumber\\
    &-\frac{l^2+l+4}{l+1} \ddot{M}^{iL}_A G^L_A\nonumber\\
    &+\frac{l}{l+1} J^L_A H^{iL}_A -\frac{4(l+1)}{(l+2)^2} \epsilon^{ijk} J^{jL}_A \dot{G}^{kL}_A \nonumber\\
    & - \frac{4}{l+2} \epsilon^{ijk} \dot{J}^{jL}_A G^{kL}_A \bigg] \bigg\} + O\left(c^{-4}\right)\,,\label{eq:dipole}\\
\dot{J}^i_A =&  \sum_{l=0}^{\infty} \frac{1}{l!} \epsilon^{ijk} M^{jL}_A G^{kL}_A + O\left(c^{-2}\right)\,,
\label{eq:spin}
\end{align}
where $M_L^A(t)=M_L^A(s_A(t))$, $J_L^A(t)=J_L^A(s_A(t))$ and the same holds for the other body-frame moments.  These
equations have to be supplemented by the equations for the multipole moments with $l\ge2$, which depend on the internal
dynamics of the bodies. In the adiabatic approximation, they are given by Eq.~\eqref{eq:adiabatic}.

The orbital equation of motion,
\begin{equation}
  \ddot{z}^i_A=\mathcal{F}^i_A[z^j_B,{\dot z}^j_B,M^L_B,{\dot M}^L_B,{\ddot M}^L_B,J^L_B,{\dot J}^L_B]\,,\label{eq:eomorbit}
\end{equation}
can be obtained from the condition $\ddot{M}^i_A=0$, which follows from the gauge condition ${M}^i_A=0$. Using
the relations between multipoles in different frames [Appendix~\ref{app:momtransf},
Eqs.~\eqref{eq:G}--\eqref{eq:Lambdaz}], Eq.~\eqref{eq:dipole} yields the explicit form of Eq.~\eqref{eq:eomorbit}.

In the case of a binary system ($N=2$), the dynamics in the center-of-mass (COM) frame is described by the
orbital separation $z^i=z^i_2-z^i_1$. If we denote the two velocities $v_A^i={\dot z}_A^i$, the relative velocity is
$v^i=v^i_2-v^i_1$. We also define the radial separation $r=|\boldsymbol{z}|=\sqrt{\delta^{ij}z^iz^j}$, and the unit vector 
$n^i=z^i/r$. The radial velocity is $\dot r=v^in^i$. The equation of motion of the orbital separation has the form
\begin{equation}
{\ddot z}^i={\ddot z}_2^i-{\ddot z}^i_1=\mathcal{F}^i_2-\mathcal{F}^i_1\,.\label{eq:eom_com}
\end{equation}
If the orbit is circular (as expected in the late inspiral)
one gets $\dot r=O(c^{-4})$. 
In this case, Eq.~\eqref{eq:eom_com} yields the radius-frequency relation $r(\omega)$, where $\omega/(2\pi)$
is the orbital frequency. 

\subsection{Lagrangian formulation and gravitational waveform}
The equations of motion~\eqref{eq:eom_com}, together with the equations for the multipole moments in the adiabatic
approximation, Eq.~\eqref{eq:adiabatic}, can be derived from a generalized action principle, in terms of a Lagrangian
function $\mathcal{L}(z^i,{\dot z}^i,{\ddot z}^i,M_A^L,{\dot M}_A^L,J_A^L)$. Therefore, given the explicit expression of
the equations of motion, it is possible to derive the corresponding Lagrangian. The binding energy of the two-body
system can be then obtained using the standard techniques of Lagrangian and Hamiltonian mechanics.

For a circular orbit, at $1$PN order the energy is a function of the radial distance $r$ only.
Replacing the radius-frequency relation $r(\omega)$, it is possible to express the energy as
a function of the orbital angular velocity $\omega$.

The GW flux emitted by the system is due to the presence of time-varying multipole moments, and is given
(up to next-to-leading order in the PN expansion) by~\cite{Blanchet:2006zz} 
\begin{align}
  F =& \frac{1}{5c^5} \dddot{M}^{ij}_{sys}\dddot{M}^{ij}_{sys} + \frac{1}{189 c^7} \ddddot{M}^{ijk}_{sys} \ddddot{M}^{ijk}_{sys}
  \nonumber\\
  &+ \frac{16}{45c^7} \dddot{J}^{ij}_{sys} \dddot{J}^{ij}_{sys} + F_{tail} +O(c^{-9})\,,\label{eq:flux}
\end{align}
where
\begin{equation}
F_{tail}=\frac{2}{5c^8}\dddot{M}^{ij}_{sys}\dot{U}^{ij}_{tail}
\end{equation}
is the tail contribution to the flux, entering at $1.5$PN order beyond the leading term (see,
e.g.,~\cite{Blanchet:2006zz,Maggiore}). The explicit expression of $U^{ij}_{tail}$ for the two-body system is given in
Eq.~\eqref{eq:tail}.  Using Eqs.~\eqref{sysS}--\eqref{sysZ} and the equations in Appendix~\ref{app:momtransf}, it is then
possible to compute the flux in terms of the mass and current moments of the bodies of the system, $M_A^L$ and $J_A^L$.

In the adiabatic approximation, the orbital and internal energy are related to the GW flux through the energy balance relation
\begin{equation}
\dot{E}=-F\,.\label{eq:enbal}
\end{equation}
Assuming Eq.~\eqref{eq:enbal} and the stationary phase approximation, the Fourier transform of the gravitational
waveform can be written as $h={\cal A}e^{{\rm i}\psi}$ where the phase $\psi(\omega)$ is given in terms of the flux and
the energy $E(\omega)$ by the differential relation
\begin{equation}
\frac{d^2\psi}{d\omega^2}=-\frac{2}{F}\frac{dE}{d\omega}\,, \label{eq:wav}
\end{equation}
(see, e.g.,~\cite{Vines:2011ud} and references therein).

\section{Tidal interactions of a spinning binary system}\label{sec:truncation}
In~\cite{Vines:2013prd,Vines:2011ud}, the approach described in the previous section has been applied to the so-called
``$M_1-M_2-J_2-Q_2$'' truncation. This is a system composed by two bodies, the first characterized by its monopole mass
moment $M_1$ only, the second characterized by its monopole mass moment $M_2$, its dipole current moment (i.e., its
spin) $J_2^i$, and its quadrupole mass moment $M_2^{ij}$, which we call $Q^{ij}$; all other multipole moments
identically vanish.  Assuming the adiabatic approximation~\eqref{eq:adiabatic} (which in this truncation reduces to
$Q^{ab}=\lambda_{2} G_2^{ab}$) and neglecting PN orders larger than one (i.e., neglecting $O(c^{-4})$ terms), the
$M_1-M_2-J_2-Q_2$ truncation describes a binary system of two nonspinning bodies with masses $M_1$ and $M_2$ and
(electric, quadrupolar) TLN $\lambda_{2}$. Indeed, in this approximation it turns out that the
spin $J^i_2$ is constant and can be consistently set to zero. Moreover, the quadrupole moment $M_1^{ij}$ tidally induced
by body $2$ (which is set to zero in the $M_1-M_2-J_2-Q_2$ truncation) can be easily derived {\it a posteriori} by
exchanging the indices $1\leftrightarrow2$, as explained below.

The PN waveform derived in~\cite{Vines:2013prd,Vines:2011ud} includes the tidal contribution to the phase up to
next-to-leading order, i.e. to overall $6$PN order.\footnote{We remark that, although the current PN waveforms of compact
  binaries only include up to $3.5$PN-order terms in the point-particle phase, the tidal interaction --~which appears at
  $5$PN order~-- cannot be neglected (and indeed is
  detectable~\cite{TheLIGOScientific:2017qsa,Annala:2017llu,EoS-GW170817}). This is due to the fact that a new
  dimensionful scale --~the NS radius $R$~-- appears in the tidal interaction, and the tidal terms in the GW phase are
  magnified by a factor $\sim(c^2R/M)^5$~\cite{Mora:2003wt}.} The quadrupolar magnetic contribution to the waveform of
a nonspinning, tidally interacting binary system, which also appears at $6$PN order, has been derived
in Refs.~\cite{Yagi:2013sva,Banihashemi:2018xfb}.

In this section, we extend the results of~\cite{Vines:2013prd,Vines:2011ud,Yagi:2013sva} by including the effects of
spin. To this aim, we apply the approach described in Sec.~\ref{sec:pntidal} to a larger truncation in which body $1$ is
characterized by its mass $M_1$ and spin $J_1^i$, while body $2$ is characterized by its mass $M_2$, its spin $J_2^i$,
its mass quadrupole moment $Q^{ij}=M^{ij}_2$, its current quadrupole moment $S^{ij}=J_2^{ij}$, its mass octupole moment
$Q^{ijk}=M_2^{ijk}$, and its current octupole moment $S^{ijk}=J_2^{ijk}$. Moreover, we neglect the terms quadratic in
the spin.  We remark that while the (mass and current) multipole moments with $l\ge2$ are assumed to be induced by tidal
interactions only\footnote{The spin-induced quadrupole moment, anyway, would only enter at quadratic order in the
  spin.}, the masses and spins $M_A$, $J^i_A$ are {\it a priori} features of the system. Therefore, in our derivation we
set to zero the $l\ge2$ moments of body $1$, but include its mass and spin. At the end of the computation, the tidally
induced ($l\ge2$) moments of body $1$ will be easily obtained by simply exchanging the indices $A=1,2$ of the two
bodies, as in~\cite{Vines:2013prd,Vines:2011ud}.
\ \\
\subsection{Equations of motion}\label{subsec:eom}
With our truncation, the equations of motion of the mass monopole (i.e., the mass), the mass dipole and the current
dipole (i.e., the spin), namely Eqs.~\eqref{eq:monopole}, \eqref{eq:dipole}, and \eqref{eq:spin}, respectively, reduce to

\begin{align}
\dot{M}_1  = &  \, O\left(c^{-4}\right) \,,\nonumber\\
  \dot{M}_2  = & -\frac{1}{c^2} \left(\frac{3}{2} Q^{ij} \dot{G}^{ij}_2 +  \dot{Q}^{ij} G^{ij}_2 \right.\nonumber\\
  &\left.+ \frac{2}{3} Q^{ijk} \dot{G}_{2}^{ijk} + \frac{1}{2}  \dot{Q}^{ijk} G_{2}^{ijk} \right) + O\left(c^{-4}\right) \,,
  \label{eq:monopole_tr}
  \end{align}
    \begin{widetext}
  \begin{align}
\ddot{M}^i_1  = & \,  M_1 \, G^i_1 + \frac{1}{c^2} \bigg[ \frac{1}{2} J^j_1 H^{ij}_1 - \epsilon^{ijk} J^{j}_1 \dot{G}^{k}_1- 2 \epsilon^{ijk} \dot{J}^{j}_1 G^{k}_1  \bigg]+ O\left(c^{-4}\right)\,,\nonumber\\
  \ddot{M}^i_2  = & \,  M_2 G^{i}_2+ \frac{1}{2} Q^{jk} G^{ijk}_2 + \frac{1}{6} Q^{jka} G^{ijka}_2+ \frac{1}{c^2} \bigg[ \frac{1}{3} \epsilon^{ijk} Q^{ja} \dot{H}^{ka}_2 + \frac{1}{2} \epsilon^{ijk} \dot{Q}^{ja} H^{ka}_2+ \frac{1}{4} \epsilon^{ijk} Q^{jab} \dot{H}^{kab}_2 + \frac{1}{3} \epsilon^{ijk} \dot{Q}^{jab} H^{kab}_2\nonumber \\
    & - 3 Q^{ij} \ddot{G}^j_2 - 6 \dot{Q}^{ij} \dot{G}^j_2-3 \ddot{Q}^{ij} G^j_2 - \frac{80}{21} Q^{ijk} \ddot{G}^{jk}_2 - \frac{65}{9} \dot{Q}^{ijk} \dot{G}^{jk}_2- \frac{10}{3} \ddot{Q}^{ijk} G^{jk}_2+\frac{1}{2} J^j_2 H^{ij}_2+\frac{1}{3} S^{jk} H^{ijk}_2 +\frac{1}{8} S^{jka} H^{ijka}_2 \nonumber \\
    &- \epsilon^{ijk} J^{j}_2 \dot{G}^{k}_2-\frac{8}{9} \epsilon^{ijk} S^{ja} \dot{G}^{ka}_2 - \frac{3}{8}\epsilon^{ijk} S^{jab} \dot{G}^{kab}_2  - 2 \epsilon^{ijk} \dot{J}^{j}_2 G^{k}_2 - \frac{4}{3} \epsilon^{ijk} \dot{S}^{ja} G^{ka}_2 - \frac{1}{2} \epsilon^{ijk} \dot{S}^{jab} G^{kab}_2 \bigg]  + O\left(c^{-4}\right)\,,\label{eq:dipole_tr}
      \end{align}
      \end{widetext}
\begin{align}
\dot{J}^i_1 = & \,  O\left(c^{-2}\right) \,, \nonumber\\  
\dot{J}^i_2 = & \,\epsilon^{ijk} Q^{ja} G^{ka}_2 + \epsilon^{ijk} Q^{jab} G^{kab}_2 + O\left(c^{-2}\right)\,.\label{eq:spin_tr}
\end{align}
The $l\ge2$ multipole moments are given, in the adiabatic approximation, by the algebraic relations
[see Eq.~\eqref{eq:adiabatic}]
\begin{equation}
\begin{split}
Q^{ab}  &=   \lambda_{2} G_2^{ab}  + \frac{\lambda_{23}}{c^2}  J_2^c H_2^{abc} \\
Q^{abc} &=   \lambda_{3} G_2^{abc} + \frac{\lambda_{32}}{c^2}  J_2^{\langle c} H_2^{ab \rangle} \\
S^{ab}  &=  \frac{\sigma_{2}}{c^2} H_2^{ab}   + {\sigma_{23}}   J_2^c G_2^{abc}  \\
S^{abc} &=   \frac{\sigma_{3}}{c^2} H_2^{abc}  + {\sigma_{32}}   J_2^{\langle c} G_2^{ab \rangle} \,.
\end{split}
\label{eq:adiabatic2}
\end{equation}
As discussed in Sec.~\ref{subsec:eom2}, the orbital equations of motion can be obtained by replacing
Eq.~\eqref{eq:dipole_tr} in the condition ${\ddot M}^i_A=0$, which is a consequence of the gauge condition $M^i_A=0$.

At $0$PN (i.e., Newtonian) order, the mass monopoles are conserved and, setting to zero the right-hand side of
Eq.~\eqref{eq:dipole_tr}, we get
\begin{align}
&M_1  G^i_1 =O(c^{-2})\nonumber\\
&M_2 G^{i}_2+ \frac{1}{2} Q^{jk} G^{ijk}_2 + \frac{1}{6} Q^{jka} G^{ijka}_2=O(c^{-2})\,,\label{eq:dipole_tr0PN}
\end{align}
where $G^i_A=G^i_{g,A}-{\ddot z}^i_A$ [Eq.~\eqref{eq:Ga}], and the global electric tidal moments are [see
Eq.~\eqref{eq:Gg}]\footnote{We remind that $\partial_L^{(A)}$ denotes the derivatives with respect to $z_A^i$, see Sec.~\ref{sec:notation}.}
\begin{equation}
  G^L_{g,A} = -\partial_L^{(A)}\phi^{ext}_A+O(c^{-2})\,,
\end{equation}
where $\phi^{ext}_A$ is the Newtonian potential on the body $A$ due to the other bodies,
\begin{equation}
\phi^{ext}_A=\sum_{B\neq A}\phi_B^{int}=-\sum_{B \neq A} \sum_{k=0}^{\infty} \frac{(-1)^k}{k!} M_B^K
\partial^{(A)}_{K} \frac{1}{|\boldsymbol{z}_A-\boldsymbol{z}_B|}\,.
\end{equation}
With our truncation, $\phi^{ext}_1=-M_2/r-Q^{ij}\partial_{ij}^{(1)}(1/r)+Q^{ijk}\partial_{ijk}^{(1)}(1/r)$,
$\phi^{ext}_2=-M_1/r$, and
\begin{align}
  M_1{\ddot z}^i_1=&-M_1\partial_i^{(1)}\phi^{ext}_1\nonumber\\
  =&\frac{M_1 M_2}{r^2} n^i+\frac{15 M_1}{2 r^4} Q^{jk}
  n^{\langle ijk \rangle}\nonumber\\
  &- \frac{35 M_1}{2r^5} Q^{jkm} n^{\langle ijkm \rangle}  +O(c^{-2}) \,,\label{orbeq0pn12}
\end{align}
while $M_2{\ddot z}^i_2=-M_1{\ddot z}^i_1+O(c^{-2})$.  In the COM frame, $z^i=z^i_2-z^i_1$
(note that $z^i_1=-\eta_2z^i$, $z^i_2=\eta_1z^i$), and Eq.~\eqref{orbeq0pn12} yields (see Sec.~\ref{sec:notation} for notation)
\begin{align}
  {\ddot z}^i=&-\frac{M}{r^2} n^i-\frac{15 M_1}{2 r^4} Q^{jk}  n^{\langle ijk \rangle}\nonumber\\
  &+\frac{35 M_1}{2r^5} Q^{jkm} n^{\langle ijkm \rangle}  +O(c^{-2})\,.\label{orbeq0pn}
\end{align}
Before proceeding with the derivation of the equations of motion at $1$PN order, we note that the Newtonian equations of
motion can also be obtained from a Lagrangian function
\begin{align}
  \mathcal{L}=&\sum_{A=1}^N\left(\frac{1}{2}M_A{\dot z}_A^2+\frac{1}{2}\sum_{l=0}^\infty\frac{1}{l!}M_A^LG_{g,A}^L+\mathcal{L}_A^{int}
  \right)\nonumber\\
  &+O(c^{-2})\,,
\end{align}
which in our truncation becomes
\begin{align}
  \mathcal{L}=&\frac{1}{2}M_1{\dot z}_1^2+\frac{1}{2}M_2{\dot z}_2^2+\frac{M_1M_2}{r}+
  \frac{1}{2}Q^{ij}G^{ij}_{g,2}\nonumber\\
&+\frac{1}{6}Q^{ijk}G^{ijk}_{g,2}
  +\mathcal{L}_2^{int}+O(c^{-2})\nonumber\\
  =&\frac{\mu{\dot z}^2}{2}+\frac{\mu M}{r}-U_{Q2}-U_{Q3}+  \mathcal{L}_2^{int}+O(c^{-2}) \,,\label{lagrnewt}
\end{align}
where [see Eqs.~\eqref{eq:Ggl23a} and~\eqref{eq:Ggl23b}]
\begin{align}
  U_{Q2}=&-\frac{1}{2}Q^{ij}G^{ij}_{g,2}=-\frac{3M_1}{2r^3}n^{  ij  }Q^{ij}+O(c^{-2})\,,\\
  U_{Q3}=&-\frac{1}{6}Q^{ijk}G^{ijk}_{g,2}=\frac{5M_1}{2r^4}n^{  ijk }Q^{ijk}+O(c^{-2}) 
  \label{newtonU}
\end{align}
are the quadrupolar and octupolar Newtonian gravitational potential energy, and $\mathcal{L}_2^{int}$ describes the
internal dynamics and depends on some internal degrees of freedom of body $2$, which we call $q_2^\alpha$.  The internal
energy of body $2$ is
\begin{equation}
E_2^{int}={\dot q}_2^\alpha\frac{\partial\mathcal{L}_2^{int}}{\partial{\dot q}_2^\alpha}-\mathcal{L}_2^{int}\,.\label{defEint0}
\end{equation}
Remarkably, without any assumption on the dependence of $\mathcal{L}_2^{int}$ on the variables $q_2^\alpha$, it is
possible to derive an equation for the internal energy. Indeed, the Euler-Lagrange equations for the
Lagrangian~\eqref{lagrnewt} give
\begin{align}
  \frac{d}{dt}\frac{\partial\mathcal{L}_2^{int}}{\partial{\dot q}_2^\alpha}=&
  \frac{\partial\mathcal{L}_2^{int}}{\partial{q}_2^\alpha}+\frac{1}{2}G^{ij}_{g,2}\frac{\partial Q^{ij}}{\partial q_2^\alpha}
  +\frac{1}{6}G^{ijk}_{g,2}\frac{\partial Q^{ijk}}{\partial q_2^\alpha}\,,
 \end{align}
and replacing in the time derivative of Eq.~\eqref{defEint0} yields
\begin{align}
  {\dot E}_2^{int}=&\left(\frac{1}{2}G^{ij}_{g,2}\frac{\partial Q^{ij}}{\partial q_2^\alpha}
  +\frac{1}{6}G^{ijk}_{g,2}\frac{\partial Q^{ijk}}{\partial q_2^\alpha}\right){\dot q}_2^\alpha+O(c^{-2})\nonumber\\
  =&\frac{1}{2}G^{ij}_{g,2}{\dot Q}^{ij}+\frac{1}{6}G^{ijk}_{g,2}{\dot Q}^{ijk}+O(c^{-2})\,.\label{eq:tidheat}
\end{align}
Equation~\eqref{eq:tidheat} represents the energy transferred by the tidal field to body $2$ (tidal heating). 

At $1$PN order, the mass monopole of the body $2$ is not conserved anymore. Its evolution equation,
Eq.~\eqref{eq:monopole_tr}, can be written as
\begin{equation}
  {\dot M}_2=\frac{1}{c^2}\left({\dot E}_2^{int}+3{\dot U}_{Q2}+
  4{\dot U}_{Q3}\right)+O(c^{-4})\,,\label{dotpartitioning}
\end{equation}
where ${\dot E}^{int}_2$ is given in Eq.~\eqref{eq:tidheat}. The above equation provides a way to partition the
mass $M_2$,
\begin{equation}
M_2=\null^nM_2+\frac{1}{c^2}\left({ E}_2^{int}+3{ U}_{Q2}+4U_{Q3}\right)+O(c^{-4}) \,, \label{partitioning}
\end{equation}
where $\null^nM_2$ is the conserved Newtonian mass of body $2$. As we shall discuss below, this partitioning of
$M_2$ is useful to find an action principle for the system.

In order to derive the expressions of $z^i_1(t)$ and $z^i_2(t)$ at $1$PN order, the right-hand side of
Eq.~\eqref{eq:dipole_tr} has to be expressed in terms of the body-frame multipole moments up to the same order. To this
aim, the expressions of $G^i_1$ and $G^L_2$ with $l=1,\dots,4$ are needed up to $1$PN order, while those of $H^{ij}_1$,
$H^L_2$ with $l=2,\dots,4$ are needed up to $0$PN order. These expressions are given, for the general case of $N$
tidally interacting bodies, in Appendix~\ref{app:momtransf} [Eqs.~\eqref{eq:G}--\eqref{eq:Lambdaz}]. The computation of
$G_2^L$, $H_2^L$ with $l=2,3$ for the truncated system is explicitly shown in Appendix~\ref{app:tidalpot} (the
derivation of the other tidal moments is similar). With our truncation, replacing the evolution equations for masses and
spins, Eqs.~\eqref{eq:monopole_tr} and \eqref{eq:spin_tr}, we find the orbital equations of motion in the form
\begin{align}
  M_1  {\ddot z}^i_1 = & F^i_{1,M} +  F^i_{1,J} + F^i_{1,Q2}\nonumber\\
  &+ F^i_{1,Q3} +  F^i_{1,S2}  +  F^i_{1,S3}  \,,\label{eq:orbeom12a}\\
  M_2  {\ddot z}^i_2 = & F^i_{2,M} +  F^i_{2,J} + F^i_{2,Q2}\nonumber\\
  &+ F^i_{2,Q3} +  F^i_{2,S2}  +  F^i_{2,S3}  \,,\label{eq:orbeom12b}
\end{align}
where the explicit expressions of the terms $F^i_{A,M}$, $F^i_{A,J}$, $F^i_{A,Ql}$, $F^i_{A,Sl}$ ($A=1,2$, $l=2,3$) are
given in Appendix~\ref{app:eom_tr}.

As a consistency check of Eqs.~\eqref{eq:orbeom12a}, \eqref{eq:orbeom12b}, we have computed the mass dipole of the
system $M_{sys}^i$, by applying Eq.~\eqref{sysM} to our truncation.  Computing the second time derivative of the mass
dipole, and replacing the orbital equations of motion~\eqref{eq:orbeom12a}, \eqref{eq:orbeom12b}, we found that ${\ddot
  M}^i_{sys}=0$ as expected from Eq.~\eqref{eq:dipole_tr} applied to the entire system (for which the tidal moments
vanish).

The equations of motion in the COM frame are obtained by subtracting those for the individual accelerations, and
replacing Eq.~\eqref{partitioning} (we recall that the total mass $M$ and the mass ratios $\eta_1$, $\eta_2$ are defined
in terms of the Newtonian masses $\null^nM_A$, see Sec.~\ref{sec:notation}):
\begin{align}
  a^i&={\ddot z}^i={\ddot z}^i_2-{\ddot z}^i_1\nonumber\\
  &=a^i_M+a^i_J+a^i_{Q2}+a^i_{Q3}+a^i_{S2}+a^i_{S3}\label{eq:orbitaleom}\,.
\end{align}
The mass and spin contributions are
\begin{align}
  a^i_M=&
  -\frac{M}{r^2} n^i
  -\frac{1}{c^2} \frac{M}{r^2} \bigg\{n^i \left[ (1+3 \nu) v^2 -
    \frac{3 \nu}{2} \dot{r}^2 \right. \nonumber \\
  & \left. -2(2+\nu) \frac{M}{r} \right]-2(2-\nu)\dot{r} v^i \bigg\}+O(c^{-4})\,,\label{eq:orbitaleomm}\\
  a^i_J=&
  \frac{\epsilon^{abc} J_2^c}{c^2 \eta_2 r^3} 
  \left[(3+\eta_2) v^a \delta^{bi}
    -3 (1+\eta_2) \dot{r} n^a \delta^{bi} + 6 n^{ai} v^b\right] \nonumber \\
    & +\frac{\epsilon^{abc} J_1^c}{c^2 \eta_1 r^3} [(3+\eta_1) v^a \delta^{bi} -3
    (1+\eta_1) \dot{r} n^a \delta^{bi} + 6 n^{ai} v^b]\nonumber\\
  &+O(c^{-4})\,.\label{eq:orbitaleomj}
\end{align}
The mass quadrupole contribution is
\begin{widetext}
\begin{align}
    a^i_{Q2}=&
     -\frac{3 Q^{ab}}{2 \eta_2 r^4} [5 n^{abi}-2n^a \delta^{bi}]+ 
    \frac{1}{c^2} \bigg\{ \frac{Q^{ab}}{r^4} \bigg[ n^{abi}
      \left(-\frac{15}{2 \eta_2} (1+3\nu) v^2+ \frac{105 \eta_1}{4} \dot{r}^2 + \frac{12}{\eta_2} (5-2\eta_2^2) \frac{M}{r} \right)   \nonumber\\
    &  +n^a \delta^{bi} \left(\frac{3}{\eta_2}(2+2\eta_2-3\eta_2^2) v^2 -\frac{15}{2\eta_2}(2-\eta_2-\eta_2^2) \dot{r}^2 -\frac{3}{\eta_2} (8-\eta_2-3\eta_2^2) \frac{M}{r} \right)
      + \frac{15}{\eta_2} (2-\nu) \dot{r} n^{ab} v^i\nonumber\\ 
      & -\frac{3}{2\eta_2}  (7-2\eta_2 + 3 \eta_2^2) n^a v^{bi} -\frac{15 \eta_1}{2 \eta_2} (1+\eta_2) \dot{r} n^{ai} v^b 
      + \frac{3 \eta_1}{2 \eta_2} v^{ab} n^i + \frac{3 }{2 \eta_2} (5-4\eta_2-\eta_2^2) \dot{r} v^a \delta^{bi} \bigg] \nonumber\\
   &  + \frac{\dot{Q}^{ab}}{r^3} [-\frac{3}{2\eta_2} (4-\eta_2) n^{ab} v^i  -\frac{15 \eta_1}{2} \dot{r} n^{abi}
      +\frac{6}{\eta_2} n^{ai} v^b 
      - \frac{3 \eta_1}{\eta_2} v^a \delta^{bi} + \frac{3}{\eta_2} (1-2\eta_2-\eta_2^2) \dot{r} n^a \delta^{bi}] \nonumber\\
   & + \frac{\ddot{Q}^{ab}}{r^2} [\frac{3}{4}n^{abi}+
      \frac{3}{2} n^a \delta^{bi}]- \frac{E_2^{int}}{r^2}n^i \bigg \}\nonumber\\
    & + \frac{3 \eta_1}{c^2 M \eta_2^2} \epsilon^{icd} J_2^c \bigg\{ \frac{Q^{ab}}{r^5}
    \bigg[ \frac{5}{2} n^{ab} (7 \dot{r} n^{d} - v^d)
      +(\delta^{ad} -5 n^{ad}) v^b -5 \dot{r} \delta^{ad} n^{b} \bigg] 
    + \frac{\dot{Q}^{ab}}{r^4} (\delta^{ad}-\frac{5}{2}n^{ad}) n^b \bigg\} \nonumber\\
    &+\frac{3 }{c^2 M \eta_1} \epsilon^{icd} J_1^c \bigg\{ \frac{Q^{ab}}{r^5}
    \bigg[ \frac{5}{2} n^{ab} (7 \dot{r} n^{d} - v^d)
      +(\delta^{ad} -5 n^{ad}) v^b -5 \dot{r} \delta^{ad} n^{b} \bigg] +
    \frac{\dot{Q}^{ab}}{r^4} (\delta^{ad}-\frac{5}{2}n^{ad}) n^b \bigg\}  \nonumber\\
  &  + \frac{3 \epsilon^{cde} J_1^a}{c^2 M \nu}
    \bigg\{ \frac{Q^{bd}}{r^5} \bigg[ 5 n^{ab} \left( \delta^{ic} v^e - \delta^{ie} v^c \right) 
      + 5 n^{ac} \left( \delta^{ib} v^e - \delta^{ie} v^b\right) +
      5 n^{bc} \left( \delta^{ia} v^e - \delta^{ie} v^a \right)
      + 35 n^{abc} \left( \dot{r}\delta^{ie}  -n^i \right) \nonumber\\
      & +\delta^{ab} \left(\delta^{ie} v^c -\delta^{ic}  \right)
      + \delta^{ac} \left(\delta^{ie} v^b -\delta^{ib}  \right)
      + 5 \left(\delta^{ab} n^c + \delta^{ac} n^b \right) \left(n^i -\dot{r}\delta^{ie}  \right) \bigg]
    - \frac{ \dot{Q}^{bd}}{r^4} \delta^{ie} \left(5 n^{abc}-\delta^{ab} n^c -\delta^{ac} n^b \right) \bigg\}\nonumber\\
&+O(c^{-4})    \,. \label{eq:orbitaleomq}
\end{align}
The mass octupole contribution is
\begin{equation}
\label{eq:massoctu}
a^i_{Q3} = \frac{5Q^{abc}}{2 \eta_2 r^5} \left(7 n^{iabc}- 3 \delta^{ic} n^{ab} \right)+O(c^{-2}) \,. 
\end{equation}
The current quadrupole contribution is
\begin{align}
\label{eq:quadcurr}
a^i_{S2} =  &  \frac{4  \epsilon^{bcd}}{c^2 \eta_2 }
\bigg\{ \frac{S^{ad}}{r^4} \bigg[ n^a \left(\delta^{ib} v^c -\delta^{ic} v^b \right) +
  n^b \left(\delta^{ia} v^c -\delta^{ic} v^a \right) + 5 n^{ab}
  \left(\dot{r}\delta^{ic} -n^i v^c \right) \bigg]
- \frac{ \dot{S}^{ad}}{r^3} \delta^{ic} n^{ab}  \bigg\}\nonumber \\
& + \frac{ J_1^c}{c^2 M \nu} \frac{S^{ab}}{r^5}
\left[ 4 \delta^{bc} \left( 5 n^{ia} -\delta^{ia} \right) + 10 \left( \delta^{ia} n^{bc} +
  \delta^{ib} n^{ac} + \delta^{ic} n^{ab}   \right)- 70 n^{iabc} \right]+O(c^{-4}) \,.
\end{align}
Finally, the current octupole contribution is
\begin{align}
\label{eq:finaleqf}
a^i_{S3} = & \frac{15}{c^2 \eta_2} \bigg\{ \frac{S^{bde}}{r^5}  n^{ab}  v^c
\bigg[\frac{7}{2}(\epsilon^{iae} n^{c}-\epsilon^{cae} n^{i})n^d -( \epsilon^{iad}
  \delta^{ce} + \epsilon^{icd}\delta^{ae} + \epsilon^{acd}\delta^{ie} )\bigg] -
\frac{\dot{S}^{bde}}{2r^4} \epsilon^{iad} n^{abe} \bigg\} \nonumber\\
& + \frac{45 J_1^c}{4 c^2 M \nu} \frac{S^{abc}}{r^6}
\bigg[ \delta^{cd} \left(\delta^{ia} n^b + \delta^{ib} n^a -7 n^{iab} \right)
  - \frac{7}{3} \left( \delta^{ia} n^{bcd} +\delta^{ib} n^{acd} +
  \delta^{ic} n^{abd} + \delta^{id} n^{abc} -9 n^{iabcd} \right) \bigg]\nonumber\\
&+O(c^{-4}) \,.
\end{align}
\end{widetext}
In the above equations, all the contributions to the orbital acceleration are given up to $1$PN order, with the
exception of the mass octupole contribution~\eqref{eq:massoctu}, which is given to $0$PN order only. 
This is sufficient to determine the GW phase up to $6.5$PN order. 

As we shall discuss in Sec.~\ref{subsec:counting}, in a circular, compact binary system, up to first order in the
spins (parallel to the orbital angular momentum), the tidally induced mass and current $l$-pole moments contribute to
the GW waveform to order $(2l+5/2)$PN through the ordinary TLNs and to order $(2l+1/2+2\delta_{l2})$PN through the
RTLNs, respectively.

\subsection{Lagrangian}
The orbital equation of motion in the COM frame, ${\ddot z}^i=a^i_M+a^i_J+a^i_{Q2}+a^i_{Q3}+a^i_{S2}+a^i_{S3}$
[Eq.~\eqref{eq:orbitaleom}], can be derived from an action principle~\cite{Vines:2013prd}. One first writes the most
general Lagrangian consistent with the truncation and at most linear in the spin, which will depend on a set of free
coefficients. Then, applying the Euler-Lagrange equations (Eq.~\eqref{eq:EL}, see below) to the Lagrangian, replacing
the evolution equations for $J_1^i$, $J_2^i$ and $E_2^{int}$, Eqs.~\eqref{eq:spin_tr} and \eqref{eq:tidheat}, and
comparing with the orbital equations of motion [Eq.~\eqref{eq:orbitaleom}], it is possible to find the values of the
coefficients, which will only depend on the masses of the two bodies. Following this approach we find that the
Lagrangian is
\begin{equation}
  \mathcal{L}_{orb}(\boldsymbol{z},\boldsymbol{v},\boldsymbol{a})=
  \mathcal{L}_M+\mathcal{L}_J+\mathcal{L}_{Q2}+\mathcal{L}_{Q3}+{\cal
  L}_{S2}+\mathcal{L}_{S3}\,.\label{eq:lagr_orb_trunc}
\end{equation}
The explicit expressions of the different terms in Eq.~\eqref{eq:lagr_orb_trunc} are given in
Sec.~\ref{subsec:summrel}, in Eqs.~\eqref{eq:LM}--\eqref{eq:LS3}.
Note that Eq.~\eqref{eq:lagr_orb_trunc} is a {\it generalized Lagrangian}, since it depends on the (relative)
acceleration $a^i$, together with the (relative) position and velocity; the action is stationary if the {\it generalized
  Euler-Lagrange equations} are satisfied,
\begin{equation}
\left(\frac{\partial}{\partial z^i}-\frac{d}{dt}\frac{\partial}{\partial v^i}+\frac{d^2}{dt^2}\frac{\partial}{\partial a^i}
\right)\mathcal{L}_{orb}=0\,.\label{eq:EL}
\end{equation}
As discussed in~\cite{Vines:2013prd}, a generalized Lagrangian is needed in order to obtain the spin contribution of the
orbital equation of motion, $a^i_J$, from an action principle.\footnote{The equations
  of motion (and then the Lagrangian) for a spinning two-body system depend on the spin supplementary condition
  assumed~\cite{Mikoczi:2016fiy}. Choosing a different spin supplementary condition (i.e., a different gauge), it is
  possible to make the Lagrangian independent of the acceleration.}
We remark that the mass monopole contribution to the acceleration $a^i_M$, Eq.~\eqref{eq:orbitaleomm}, is only due to
the monopole term of the Lagrangian, $\mathcal{L}_M$. The spin contribution to the acceleration $a^i_J$, Eq.~\eqref{eq:orbitaleomj}, is only due to the spin term of the Lagrangian, $\mathcal{L}_J$.  The mass quadrupole
contribution to the acceleration $a^i_{Q2}$, Eq.~\eqref{eq:orbitaleomq}, arises from terms in $\mathcal{L}_M$,
$\mathcal{L}_J$ and $\mathcal{L}_{Q2}$. Finally, the current quadrupole and the (mass and current) octupole contributions to the acceleration
$a^i_{S2}$, $ a^i_{Q3}$, $a^i_{S3}$, Eqs.~\eqref{eq:massoctu}--\eqref{eq:finaleqf}, arise from $\mathcal{L}_{S2}$, $\mathcal{L}_{Q3}$ and
$\mathcal{L}_{S3}$, respectively.

It is possible to extend the Lagrangian $\mathcal{L}_{orb}$ in order to also describe the adiabatic evolution of the
mass and current quadrupole and octupole moment ($Q^{ij}$, $S^{ij}$, $Q^{ijk}$, $S^{ijk}$), i.e., to enforce the adiabatic relations in Eq.~\eqref{eq:adiabatic}.
In this derivation, we shall use the explicit expressions of the $l=2,3$ tidal moments of body $2$, which have been
derived in Appendix~\ref{app:tidalpot} [see Eqs.~\eqref{eq:G2ab}--\eqref{eq:G21PN}].

At $0$PN order, the mass multipole contributions to the Lagrangian are
\begin{align}
  \mathcal{L}_{Q2}=&\frac{1}{2}G_2^{ab}Q^{ab}+O(c^{-2})\,,\\
  \mathcal{L}_{Q3}=&\frac{1}{6}G_2^{abc}Q^{abc}+O(c^{-2})\,,
\end{align}
while $\mathcal{L}_{S2}\sim\mathcal{L}_{S3}\sim O(c^{-2})$.\footnote{For a generic mass multipole moment of order $l$,
  the contribution to the Newtonian Lagrangian is $\mathcal{L}_{Ql}=\frac{1}{l!} G_2^L Q^L + O(c^{-2})$. In the case of
  current multipole moments, the structure is akin to the Newtonian one, but at order $1$PN,
  $\mathcal{L}_{Sl}=\frac{1}{c^2} \frac{1}{l!} \frac{l}{l+1} H_2^L S^L + O(c^{-4})$.}

Up to $1$PN order, the mass quadrupole contribution $\mathcal{L}_{Q2}$~\eqref{eq:LMQ} can be written as
\begin{equation}
\mathcal{L}_{Q2}=U^{ab}Q^{ab}+V^{ab}{\dot Q}^{ab}+WE_2^{int}+O(c^{-4})\,,\label{eq:LQ2}
\end{equation}
where $U^{ab}(\boldsymbol{z},\boldsymbol{v})$, $V^{ab}(\boldsymbol{z},\boldsymbol{v})$,
$W(\boldsymbol{z},\boldsymbol{v})$ are the coefficients appearing in Eq.~\eqref{eq:LMQ}, i.e.,
\begin{align}
  U^{ab} = & \, \frac{3\eta_1 M}{2r^3} n^{ab}  +  \frac{1}{c^2} \frac{M}{r^3}
  \bigg[ n^{ab} \bigg( \frac{3 \eta_1}{4} (3+\nu)  v^2\nonumber\\
    &+ \frac{15 \nu \eta_1}{4} \dot{r}^2  -\frac{3 \eta_1}{2} (1+3\eta_1)\frac{M}{r} \bigg)\nonumber\\
    &+\frac{3 \eta_1^2}{2} v^{ab}  -\frac{3 \eta_1^2}{2}(3+\eta_2) \dot{r} n^a v^b \bigg]  \nonumber \\
  & +  \frac{1}{c^2 } \frac{3}{r^4}\epsilon^{eca} J_1^d \left( 5 n^{bcd} -\delta^{bd} n^c -\delta^{cd} n^b\right) v^e\,,
  \label{def:U}\\
V^{ab} = & \, \frac{1}{c^2} \frac{M}{r^2} [ -\frac{3 \nu }{2}n^a v^b -\frac{3 \nu }{4} \dot{r} n^{ab}]\,, \label{def:V}  \\
W = & \, \frac{1}{c^2}  \left[\frac{\eta_1^2}{2}v^2 + \eta_1 \frac{M}{r} \right]\,.\label{def:W}
\end{align}
As discussed in Sec.~\ref{subsec:eom}, we do not explicitly compute the $O(c^{-2})$ corrections in the mass octupole
contribution because they do not affect the tidal contribution to the GW phase to $6.5$PN order. The current quadrupole
and octupole contributions to the Lagrangian, up to $1$PN order, can be written as:
\begin{align}
\mathcal{L}_{S2}=&\frac{1}{3c^2}H_2^{ab}S^{ab}+O(c^{-4})\,,\\
\mathcal{L}_{S3}=&\frac{1}{8c^2}H_2^{abc}S^{abc}+O(c^{-4})\,.
\end{align}
Comparing Eqs.~\eqref{def:U}--\eqref{def:W} with the expression of $G_2^{ab}$ at $1$PN order [Eq.~\eqref{eq:G21PN}], we find that
\begin{equation}
G_2^{ab}=2(1+W)(U^{\langle ab \rangle}-{\dot V}^{\langle ab \rangle })+O(c^{-4})\,, \label{eq:G1pn}
\end{equation} 
an expression which will be useful below. 

Let us now define the Lagrangian
\begin{align}
  \mathcal{L}(\boldsymbol{z},\boldsymbol{v},\boldsymbol{a},Q^L,\dot{Q}^L,S^L)=& 
  \mathcal{L}_{orb}(\boldsymbol{z},\boldsymbol{v},\boldsymbol{a},Q^L,\dot{Q}^L,S^L) \nonumber \\
  & +  \mathcal{L}_2^{int}(Q^L,S^L)\,.\label{eq:lagr_tot_trunc}
\end{align}
The internal Lagrangian $\mathcal{L}_2^{int}$ only depends on the internal degrees of freedom $Q^{ab}$, $Q^{abc}$,
$S^{ab}$, $S^{abc}$, and describes the internal dynamics, while the orbital Lagrangian depends both on the orbital
degrees of freedom and on the momenta $Q^L$, $S^L$. Note that $\mathcal{L}_{Q2}$ also depends on $E_2^{int}/c^2$ [see
Eqs.~\eqref{eq:LQ2}, \eqref{def:W}]; in order to write the Euler-Lagrange equations, we need to know the explicit form
of $E_2^{int}(Q^L,S^L)$, at $0$PN order. To this aim, note that, replacing the adiabatic relations~\eqref{eq:adiabatic}
in Eq.~\eqref{eq:tidheat} we find, at leading order,
\begin{align}
  {\dot E}_2^{int}=&\frac{1}{2}G_2^{ab}{\dot Q}^{ab}+\frac{1}{6}G_2^{abc}{\dot Q}^{abc}+
  O(c^{-2})\nonumber\\
  =&\frac{1}{4\lambda_2}\frac{d}{dt}(Q^{ab}Q^{ab})+\frac{1}{12\lambda_3}\frac{d}{dt}(Q^{abc}Q^{abc})\nonumber\\
&  +O(c^{-2})\label{eq:tidheat1}\,.
\end{align}
Therefore, the internal energy (at Newtonian level) has the form
\begin{align}
  E_2^{int}=&\frac{1}{4\lambda_2}Q^{ab}Q^{ab}+\frac{1}{12\lambda_3}Q^{abc}Q^{abc}+O(c^{-2})\,. \label{EintN}
\end{align}

The corresponding internal Lagrangian is $\mathcal{L}_2^{int}=-E_2^{int}$, and the
Euler-Lagrange equations
\begin{equation}
\left(\frac{\partial}{\partial q}-\frac{d}{dt}\frac{\partial}{\partial \dot{q}}\right)\mathcal{L}=0 
  \label{eq:EulLag}
\end{equation}
(where $q=\{Q^{ab}, Q^{abc},S^{ab}, S^{abc}\}$)
give the Newtonian adiabatic relations, $Q^{ab}=\lambda_2G_2^{ab}+O(c^{-2})$, $Q^{abc}=\lambda_3G_2^{abc}+O(c^{-2})$.

At $1$PN order we cannot use Eq.~\eqref{eq:tidheat1} to derive the expression of the internal energy, because that
equation is only given up to Newtonian order. We instead look for an expression which reduces to~\eqref{EintN} at $0$PN
order, and which yields the correct adiabatic relations~\eqref{eq:adiabatic} at $1$PN order. We find that
if\footnote{Note that this is the most general Lagrangian function, which can be built from the multipole moments of
  our truncation, and which is at most quadratic in the internal degrees of freedom and linear in the spin.}
\enlargethispage{4.4cm}
\begin{align}
  \mathcal{L}_2^{int}=&-E_2^{int}=-\frac{1}{4\lambda_2}Q^{ab}Q^{ab}-\frac{1}{12\lambda_3}Q^{abc}Q^{abc}\nonumber\\
  &-\frac{1}{6\sigma_2}S^{ab}S^{ab}-\frac{1}{16\sigma_3}S^{abc}S^{abc}\nonumber\\
  &+\alpha J_2^a Q^{bc}S^{abc}+ \beta J_2^a S^{bc}Q^{abc}\,,\label{EintPN}
\end{align}
the Euler-Lagrange equations~\eqref{eq:EulLag} give 
\begin{align}
  Q^{ab}  =  &  \lambda_2G_2^{ab}+ 2\lambda_{2} \alpha   J_2^c S^{abc}+O(c^{-4})\,,\nonumber\\
  Q^{abc} =  & \lambda_{3} \left(G_2^{abc}+O(c^{-2})\right) +6\lambda_3 \beta
  J_2^{\langle c} S^{ab \rangle}\nonumber\\
  &+O(c^{-4}) \,,\nonumber\\
  S^{ab}  =  & \frac{\sigma_{2}}{c^2} H_{2}^{ab}   + 3\sigma_2 \beta  J_2^c Q^{abc} +O(c^{-4})\,, \nonumber\\
  S^{abc} =  & \frac{\sigma_{3}}{c^2} H_{2}^{abc}  + 8\sigma_3 \alpha
  J_2^{\langle c} Q^{ab \rangle}+O(c^{-4})\,,
\end{align}
where $G_2^{ab}$ in the above expression is precisely given by Eq.~\eqref{eq:G1pn} to this PN order.

In order to simplify the above expressions, it is useful to substitute the adiabatic expressions for $Q^L$ and $S^L$ to
lowest order in the spin, and truncate the result to linear order in the spin. This is clearly consistent with our
perturbative scheme which neglects quadratic and higher-order spin terms. This substitution yields
\begin{align}
  Q^{ab}  =  &  \lambda_2G_2^{ab}+ \frac{2\lambda_{2}\sigma_3 \alpha}{c^2}  J_2^c H_{2}^{abc}+O(c^{-4})\,,\nonumber\\
  Q^{abc} =  & \lambda_{3} \left(G_{2}^{abc}+O(c^{-2})\right) +\frac{6\lambda_3\sigma_2 \beta}{c^2}
  J_2^{\langle c} H_{2}^{ab \rangle}\nonumber\\
  &+O(c^{-4}) \,,\nonumber\\
  S^{ab}  =  & \frac{\sigma_{2}}{c^2} H_{2}^{ab}   + 3\lambda_3\sigma_2 \beta  J_2^c G_{2}^{abc} +O(c^{-4})\,, \nonumber\\
  S^{abc} =  & \frac{\sigma_{3}}{c^2} H_{2}^{abc}  + 8\lambda_2 \sigma_3 \alpha
  J_2^{\langle c} G_{2}^{ab \rangle}+O(c^{-4})\,,\label{eq:adiabatic1}
\end{align}
which coincide with the adiabatic relations~\eqref{eq:adiabatic} [with the replacement~\eqref{rotTLNs}] at $1$PN order,
to leading order in the tidal moments, and to linear order in the spin.  We remark that Eq.~\eqref{EintPN} has been
obtained under the assumption that the multipole moments are tidally induced and neglecting the contributions
  quadratic in the spin; therefore, in the Newtonian limit Eq.~\eqref{EintPN} reduces to Eq.~\eqref{EintN}.

Finally, we note that, replacing the adiabatic relations Eqs.~\eqref{eq:adiabatic} in Eq.~\eqref{eq:spin_tr}, it
follows that the spins of the two bodies are constant.
\subsection{Gravitational waveform}\label{subsec:waveform}
In order to derive the gravitational waveform, we need the radius-frequency relation (to $1$PN order) for a circular
binary. In this case, $n^i=(\cos(\omega t),\sin(\omega t),0)$, $v^i=r\omega\phi^i$ where $\phi^i=(-\sin(\omega
t),\cos(\omega t),0)$. We assume the spins of the two bodies to be parallel to the orbital angular momentum.  Therefore,
we can write $J^i_A=J_As^i$ ($A=1,2$), where $s^i=\epsilon^{ijk}n^j\phi^k=(0,0,1)$, and define the dimensionless spin
variables $\chi_A = c J_A/(\eta_A M)^2$.  Replacing these expressions in the orbital equations of motion,
Eq.~\eqref{eq:orbitaleom}, using the Newtonian orbital acceleration Eq~\eqref{orbeq0pn} to simplify the expressions of
higher PN order, and imposing the adiabatic relations~\eqref{eq:adiabatic}, we find
\begin{widetext}
\begin{align}
  r =&  \frac{M^{1/3}}{ \omega^{2/3}} \bigg{\{}  1  + \frac{\nu-3}{3} x +
  \left[ \frac{(\eta_1-3)\eta_1}{3} \chi_1 + \frac{(\eta_2-3) \eta_2}{3} \chi_2 \right]x^{1.5}+  O(x^2) +\frac{3 \eta_1}{\eta_2}
  \frac{c^{10}}{M^5} \lambda_2 x^5  \nonumber\\
    &+ \left[-\frac{\eta_1}{2\eta_2} \left(6- 26
    \eta_2 + \eta_2^2 \right) \frac{c^{10}}{M^5} \lambda_2   + \frac{16 \eta_1}{\eta_2} \frac{c^8}{M^5} \sigma_2 \right]x^6+ \left[ \left( 2\eta_1 \left(9 -2 \eta_2 \right)
    \chi_2 -\frac{4 \eta_1^3}{\eta_2}
    \chi_1 \right) \frac{c^{10}}{M^5} \lambda_2- \frac{48 \eta_1^2}{\eta_2} \chi_1
    \frac{c^8}{M^5} \sigma_2 \right.\nonumber\\
&\left. + \frac{\nu \chi_2 c^{10}}{M^4} \left( - 24   \lambda_{23}  
+ 8   \lambda_{32} + 8   \sigma_{23} - 6   \sigma_{32} \right) \right] x^{6.5}
  +O(x^7) \bigg{\}} \,,
  \label{eq:romega}
\end{align}
where the first line (up to order $O(x^2)$) refers to the point-particle terms, and the others refer to the tidal terms to linear order in
the spin. We recall that $x=(\omega M)^{2/3}/c^2=v^2/c^2+O(c^{-4})$.

Replacing the adiabatic relations~\eqref{eq:adiabatic} in the Lagrangian~\eqref{eq:totallagrangian} yields the {\it
  reduced Lagrangian} 
\begin{align}
  \mathcal{L}(\boldsymbol{z},\boldsymbol{v},\boldsymbol{a}) =&
  \frac{\mu v^2}{2}+\frac{\mu M}{r}\left(1+\frac{3\eta_1}{2\eta_2}
  \frac{\lambda_2}{r^5} + \frac{15 \eta_1}{2 \eta_2} \frac{\lambda_3}{r^7}\right)+\frac{\mu}{c^2}\left\{\frac{1-3\nu}{8} v^4+\frac{M}{r}
  \left[v^2\left(\frac{3+\nu}{2}+\frac{3\eta_1^2(5+\eta_2) }{4 \eta_2}\frac{\lambda_2}{r^5}\right)\right.\right.\nonumber\\
    &\left.\left.+{\dot r}^2\left(\frac{\nu}{2} -\frac{9 \eta_1(1-6\eta_2+
      \eta_2^2)}{2 \eta_2}\frac{\lambda_2}{r^5}\right)  +\frac{M}{r} \left(-\frac{1}{2}+\frac{3 \eta_1(-7+
      5\eta_2)}{2\eta_2}\frac{\lambda_2}{r^5}\right)\right]\right\} \nonumber \\
  & +\frac{\epsilon^{abc}}{c^2}   v^b \left[ \left( \eta_2 J_1^a + \eta_1
    J_2^a \right)  \frac{2M}{r^2} n^c  + \left( \eta_2^2 J_1^a + \eta_1^2 J_2^a  \right) \frac{a^c}{2}  \right]    + \frac{\lambda_2}{c^2}\frac{9 \eta_1 M}{r^7} \epsilon^{abc} n^a J_1^b v^c \nonumber \\
  & + \frac{\sigma_2}{c^4} \left[ \frac{12 \eta_1^2 M^2}{r^6} \left(v^2-\dot{r}^2
    \right) + \frac{24 \eta_1 M}{r^7} \epsilon^{abc} n^a J_1^b v^c\right]  + \frac{\sigma_3}{c^4} \left[ \frac{60 \eta_1^2 M^2}{r^8} \left(v^2-\dot{r}^2
    \right) + \frac{180 \eta_1 M}{r^9} \epsilon^{abc} n^a J_1^b v^c\right] \nonumber \\
  & + \frac{\eta_1^2 M^2}{c^2 r^7}\left( 48 \lambda_2 \sigma_3 {\alpha}  -36 \lambda_3
  \sigma_2 {\beta}  \right) \epsilon^{abc} n^a J_2^b v^c \,.
\end{align}
Note that the contributions from the RTLNs in the above equation only enter through the terms proportional to $\alpha$
and $\beta$. However, these terms are linear in the velocity, and therefore they do not contribute to the conserved
energy below and to the GW waveform. As we show below, the RTLNs enter in the GW waveform through the radius-frequency
relation and through the GW flux.

From the above reduced Lagrangian, the conserved energy of our truncation then reads~\cite{Mikoczi:2016fiy}
\begin{align}
  E=& v^i \left(\frac{\partial\mathcal{L}}{\partial v^i} -\frac{d}{dt}\frac{\partial\mathcal{L}}{\partial a^i} \right)+ 
  a^i\frac{\partial\mathcal{L}}{\partial a^i} -\mathcal{L}\nonumber\\
  =&\frac{\mu v^2}{2}-\frac{\mu M}{r}\left(1+
  \frac{3\eta_1}{2\eta_2} \frac{\lambda_2}{r^5} +  \frac{15 \eta_1}{2 \eta_2} \frac{\lambda_3}{r^7}\right)
 +\frac{\mu}{c^2}\left\{ \frac{3(1-3\nu)}{8} v^4\right. +\frac{M}{r}\left[v^2\left(\frac{3+\nu}{2}+\frac{3\eta_1^2(5+\eta_2)}{4 \eta_2} \frac{\lambda_2}{r^5}\right)
   \right.\nonumber\\
  &+{\dot r}^2\left(\frac{\nu}{2} -\frac{9 \eta_1(1-6\eta_2+
     \eta_2^2)}{2 \eta_2}\frac{\lambda_2}{r^5}\right) \left.\left. -\frac{M}{r} \left(-\frac{1}{2}+\frac{3 \eta_1(-7+5\eta_2)}{2\eta_2}
   \frac{\lambda_2}{r^5}\right)\right]\right\} +\frac{\epsilon^{abc}}{c^2}   v^b a^c \left( \eta_2^2 J_1^a + \eta_1^2 J_2^a  \right) \nonumber \\
 & + \frac{\sigma_2}{c^4}  \frac{12 \eta_1^2 M^2}{r^6} \left(v^2-\dot{r}^2 \right)+\frac{\sigma_3}{c^4}  \frac{60 \eta_1^2 M^2}{r^8} \left(v^2-\dot{r}^2 \right) \,.\label{cons_en}
\end{align}
For a circular orbit, replacing the radius-frequency relation~\eqref{eq:romega}, the conserved energy can be written as
\begin{align}
  E =& - \frac{ \mu}{2} (M \omega)^{2/3} \Bigg{\{}  1 - \frac{9+ \nu}{12} x  +\left[ \frac{2\eta_2(\eta_2+3)}{3 }  \chi_2+\frac{2\eta_1(\eta_1+3)}{3 } \chi_1\right]x^{1.5}+O(x^2)  - \frac{9 \eta_1}{\eta_2} \frac{c^{10}}{M^5} \lambda_2 x^5  \nonumber\\
& -\left[ \frac{11 \eta_1}{2 \eta_2}\left(3+2\eta_2+3\eta_2^2 \right)  \frac{c^{10}}{M^5}
  \lambda_2 + \frac{88 \eta_1}{\eta_2} \frac{c^{8}}{M^5} \sigma_2 \right]x^6   + \left\{ \left[24 \eta_1 (\eta_2-3) \chi_2 +   \frac{24 \eta_1^3 }{\eta_2}  \chi_1
  \right] \frac{c^{10}}{M^5}  \lambda_2 \right. \left. +\frac{192\eta_1^2 }{\eta_2} \chi_1 \frac{c^{8}}{M^5} \sigma_2 \right.   \nonumber\\
& \left. + \frac{c^{10} \nu \chi_2 }{M^4} \left(96 \lambda_{23} -32 \lambda_{32} -32 \sigma_{23}
+24 \sigma_{32} \right) \right\} x^{6.5}  
+O(x^7) \Bigg{\}}\,.
\end{align}
Note that in this equation the RTLNs ($\lambda_{23}$, $\lambda_{32}$, $\sigma_{23}$, and $\sigma_{32}$) appear
explicitly, since the adiabatic relations have been used to obtain Eq.~\eqref{eq:romega}.

The GW flux (at $1.5$PN) is given in Eq.~\eqref{eq:flux}, whereas the multipole moments of system are given by
Eqs.~\eqref{sysS}--\eqref{sysZ}. Within our truncation and for a circular orbit they read\footnote{We recall that to get
  the GW phase up to $6.5$PN order we need to include the mass octupole moment $Q^{ijk}$ of the body 2 only at the
  leading order. Since this term enters to the GW flux at the next-to-leading order, we can safely neglect its
  contribution to the system multipole moments.}
\begin{align}
  M^{ij}_{sys} = & Q^{ij}+  \mu r^2 n^{\langle ij \rangle} +  \frac{1}{c^2}
  \bigg{\{}    \mu r^2 \left[ \left(  \frac{29(1-3\nu) v^2}{42}   + \frac{(8\nu -5) M }{7r} \right)  n^{\langle ij \rangle}  +
    \left( \frac{11(1-3\nu) }{21} \right) v^{\langle ij \rangle} \right] \nonumber\\
  & + \frac{4 r}{3} \left( 2 v^a n^{\langle i } - n^a v^{\langle i } \right)
  \epsilon^{j \rangle a b} \left(\eta_1^2 J_2^b + \eta_2^2 J_1^b \right)  + \bigg[ \left(E_2^{int} + 3 U_{Q2}  \right) \eta_1^2 r^2 n^{\langle ij \rangle}  -\frac{\eta_1 M}{42 r} \bigg( 2(46 \eta_1^2 + 109 \eta_1 \eta_2 +63 \eta_2^2 )Q^{ij} \nonumber \\
& -3  (52 \eta_1^2 + 4 \eta_1 \eta_2 -25 \eta_2^2 ) n^{\langle ij \rangle ab} Q^{ab}  -6 (15 \eta_1^2 + 21 \eta_1 \eta_2 +11 \eta_2^2 ) n^{a \langle i} Q^{j \rangle a}   \bigg)  + \frac{\eta_1^2 v^2}{42}  \left( 29 Q^{ij} -66 \phi^{a \langle i} Q^{ j \rangle a} \right)\nonumber\\
    & + \frac{2 \eta_1^2 r}{21} \left( n^{\langle i } \dot{Q}^{j \rangle a} v^a
    + 8 v^{\langle i } \dot{Q}^{j \rangle a} n^a \right) + \frac{\eta_1^2 r^2}{42} \left(11 \ddot{Q}^{ij}  -12 n^{a \langle i } \ddot{Q}^{j \rangle a} \right) \bigg]  +  \frac{8 \eta_1}{9} \left( 2 \epsilon^{ab \langle i } S^{j \rangle b}  v^a -  r \epsilon^{ab \langle i }
  \dot{S}^{j \rangle b}  n^a \right)\bigg{\}} \nonumber \\
  &+ O(c^{-4}) \label{eq:sysq2}
  \end{align}
  \begin{equation}
  M^{ijk}_{sys} =   \mu r^3 (\eta_1 - \eta_2)  n^{\langle ijk \rangle}
  + 3 \eta_1 r Q^{\langle ij} n^{k \rangle} + O(c^{-2})\label{eq:sysq3}
  \end{equation}
    \begin{equation}
  J^{ij}_{sys} =  S^{ij} +  \mu  r^2 (\eta_1 -\eta_2)
  \epsilon^{ab \langle i} n^{j \rangle a} v^b +\frac{3r}{2}
  \left(\eta_1 J_2^{\langle i }- \eta_2 J_1^{\langle i } \right) n^{ j \rangle}
  +\frac{\eta_1 }{2} \left(-2  \epsilon^{ ab \langle i } Q^{j \rangle b}v^a + r \epsilon^{ ab \langle i } \dot{Q}^{j \rangle b} n^a  \right) +O(c^{-2}) \,.\label{eq:syss2}
\end{equation}
The tail term appearing in Eq.~\eqref{eq:flux} is given at the leading-order by~\cite{Blanchet:2006zz,Maggiore}
\begin{equation}
  U^{ij}_{tail}(U) = 2M \int_0^{\infty} \ddddot{M}^{ij}_{sys}(U-\tau)
  \left[ \log{\left(\frac{c \tau}{2 r_0} \right)+ \frac{11}{12}} \right] d\tau \,, \label{eq:tail}
\end{equation}
where $U=t-r/c- \left( 2M/c^3 \right) \log{\left( r/r_0\right)}$ is the retarded time in radiative coordinates and $r_0$
a gauge-dependent arbitrary constant due to the freedom of choice of the radiative coordinates themselves. The final
result in the GW flux is independent of $r_0$~\cite{Blanchet:2006zz,Damour:2012yf}.  Replacing
Eqs.~\eqref{eq:sysq2}--\eqref{eq:tail} into Eq.~\eqref{eq:flux}, and using the adiabatic relations~\eqref{eq:adiabatic}
and the radius-frequency relation~\eqref{eq:romega}, the energy loss by GW emission can be written as
\begin{align}
\dot{E} =&  -\frac{32}{5} \nu^2 c^5 x^5 \Bigg{\{}  1 -\left(\frac{1247}{336}+ \frac{35}{12}\nu \right)x  + \left[4\pi- \frac{\eta_2(5+6\eta_2)}{4} \chi_2  -\frac{\eta_1(5+6\eta_1)}{4} \chi_1  \right] x^{1.5}+O(x^2)\nonumber\\
& + \frac{6(3-2 \eta_2)}{\eta_2}\frac{c^{10}}{M^5} \lambda_2 x^5  + \left[ \frac{\left(-704-1803 \eta_2+4501 \eta_2^2 -2170 \eta_2^3 \right)}{28 \eta_2}
  \frac{c^{10}}{M^5} \lambda_2  \right.  + \frac{2(113 -114\eta_2)}{3 \eta_2}  \frac{c^{8}}{M^5} \sigma_2  \Bigg] x^6 \nonumber\\
& + \left\{ \left[  \frac{24\pi(3-2 \eta_2)}{\eta_2} +  \frac{(667- 939 \eta_2 + 304 \eta_2^2)}{8}
  \chi_2  \right. \right. + \left. \frac{(-395 +1110 \eta_2 -1019 \eta_2^2 + 304 \eta_2^3)}{ 8 \eta_2}   \chi_1 \right]
\frac{c^{10}}{M^5} \lambda_2 \nonumber\\ 
& + \left[ \chi_2   +\frac{(-613+1225
    \eta_2-612 \eta_2^2)}{3 \eta_2}  \chi_1 \right] \frac{c^8}{M^5} \sigma_2   \nonumber\\
& + \frac{c^{10} \chi_2}{M^4} \bigg[ 8\eta_2(-17+12\eta_2)  \lambda_{23}   + 32 \nu  \lambda_{32}   \left. \left. +\frac{ \eta_2 (113-114 \eta_2)}{3}   \sigma_{23}   - 24 \nu  \sigma_{32}
  \right] \right\} x^{6.5}+O(x^7) \Bigg{\}}\,.
\end{align}

Finally,  Eq.~\eqref{eq:wav} gives the phase of gravitational waveform:
\begin{align}
  \psi(x)  = & \frac{3}{128 \nu x^{5/2}} \Bigg\{ 1 + \left(\frac{3715}{756}+\frac{55}{9} \nu \right) x +
  \left(\frac{113}{3}(\eta_1 \chi_1 + \eta_2 \chi_2)  -\frac{38}{3} \nu (\chi_1 + \chi_2 )  \right.  -16\pi \bigg) x^{1.5}+O(x^2)\nonumber \\
  & + \left(264 -\frac{288}{\eta_2}\right) \frac{c^{10}\lambda_2}{M^5}  x^5  + \left[ \left(  \frac{4595}{28}- \frac{15895}{28 \eta_2}  + \frac{5715 \eta_2}{14} -
    \frac{325 \eta_2^2}{7}   \right)\frac{c^{10}\lambda_2}{M^5} + \left( \frac{6920}{7} - \frac{20740}{21 \eta_2} \right) \frac{c^{8}\sigma_2}{M^5} \right] x^6 \nonumber\\
  &+  \left\{ \left[ \left( \frac{593}{4} - \frac{1105}{8 \eta_2} +\frac{567 \eta_2}{8}
    -81 \eta_2^2 \right) \chi_1\right. \right. \left. + \left( -\frac{6607}{8} +\frac{6639 \eta_2}{8} -81 \eta_2^2 \right) \chi_2 \right.  \left. -  \pi\left(264 -\frac{288}{\eta_2}\right) \right] \frac{c^{10}\lambda_2}{M^5} \nonumber\\
  &  + \left[ \left(-\frac{9865}{3} + \frac{4933}{3 \eta_2} + 1644 \eta_2 \right) \chi_1
    -\chi_2 \right]   \frac{c^{8}\sigma_2}{M^5}  +\frac{c^{10} \chi_2}{M^4}\bigg[ \left(   856 \eta_2 -  816 \eta_2^2 \right){\lambda_{23}}   - \left(\frac{833 \eta_2}{3} - 278 \eta_2^2  \right) {\sigma_{23}} \nonumber\\
    &- \nu \left(272 {\lambda_{32}} -204 {\sigma_{32} }\right)  \bigg] \bigg\} x^{6.5}+O(x^7) \Bigg\} \,.\label{phase0}
\end{align} 
\end{widetext}
As previously explained, to obtain the full GW phase up to octupole mass and current moments for both bodies, it is
sufficient to add to Eq.~\eqref{phase0} the same expression obtained by exchanging the indices $1$ and $2$ of the two
bodies.
The result is given in Eq.~\eqref{PHASE}. 
\subsection{PN order counting of the spin-tidal terms}\label{subsec:counting}
As shown in Sec.~\ref{subsec:waveform}, the spin-tidal couplings computed above modify the GW phase~\eqref{PHASE} at
$6.5$PN order, i.e., $1.5$PN order after the leading-order (electric, quadrupolar) tidal deformability term, and $0.5$PN
order \emph{before} the standard, electric, octupolar tidal deformability term. It is interesting to generalize this
counting to multipole moments and tidal moments of generic harmonic index $l$.

Let us start by considering the contribution from RTLNs.
First of all, we notice that $Q^L$ (respectively, $S^L$) enters the waveform at $l$PN (respectively, $(l+1/2)$PN)
order~\cite{Poisson:1997ha,Mikoczi:2005dn,Blanchet:2006zz}. Indeed, the contribution of $Q^L$ to the radial acceleration
in the binary system is of the order~\cite{Racine:2004hs}
\begin{equation}
 |a^i|\sim\frac{Q^L}{r^{l+2}}\,, 
\end{equation}
to be compared to the Newtonian term $|a^i|\sim M/r^2$. On the other hand, the contribution of $S^L$ is suppressed\footnote{Since $G^L$ and $H^L$ enter, respectively, at $l+1$ and $l+3/2$ leading PN order, the leading-order corrections from the ordinary TLNs in the nonspinning case is ($2l+1$)PN and ($2l+2$)PN for electric and magnetic TLNs of order $l$, respectively. \label{footnotePN}} by an
extra power of $v/c$.
Furthermore, according to the selection rules discussed in Ref.~\cite{Pani:2015nua}, $Q^L$ (respectively, $S^L$) is induced by $H^{L\pm1}$
(respectively, $G^{L\pm1}$) at linear order in the spin.
Since $H^{L\pm1}\sim v G^{L\pm1}\sim v/r^{l+1\pm1}\sim (l+3/2\pm1)$PN, we obtain that the PN order of the corrections
proportional to the spin and to the RTLNs is
\begin{equation}
 {\rm PN~order}_{\rm RTLNs}=l+\left(l+\frac{3}{2}\pm1\right)=2l+\frac{3}{2}\pm1\,, \label{PNorderRTLNs}
\end{equation}
where the upper and lower signs refer to the coupling between an $l$-pole moment and the tidal moment with $l+1$ and $l-1$, respectively. 
This result is interesting for the following reasons:
\begin{itemize}
 \item[(i)] When $l\geq3$, the lower sign clearly provides the lowest PN correction, namely $(2l+1/2)$PN. For example, the
   coupling between $l=4$ multipole moments with octupolar tidal moments would give rise to $8.5$PN terms, whereas for
   $l=3$ we obtain the $6.5$PN correction computed in the previous sections.
 \item[(ii)] On the other hand, for $l=2$ the absence of any dipolar tidal moment that could potentially induce a quadrupole
   moment imposes to use the upper sign in the above equation. This gives again a $6.5$PN term, consistent with our
   analysis.
 \item[(iii)] When compared to the PN order of the usual TLNs in the nonspinning case (namely ($2l+1$)PN and ($2l+2$)PN for
   electric and magnetic TLNs of order $l$, respectively, see footnote~\ref{footnotePN}), it is clear that the
   contribution in Eq.~\eqref{PNorderRTLNs} with the lower sign enters at \emph{lower} PN order than the usual TLNs in the nonspinning case for any $l\geq3$. Indeed, for any $l\geq3$, it
   enters at $0.5$PN ($1.5$PN) \emph{before} the electric (magnetic) TLN of order $l$.
 \item[(iv)] For both signs in Eq.~\eqref{PNorderRTLNs}, the PN order of RTLNs is the \emph{average} between the PN order of
   an ordinary tidal term of order $l$ and the tidal term of opposite parity and with $l\pm1$. This is reminiscent of
   the selection rules discussed in Ref.~\cite{Pani:2015nua}.
\end{itemize}

Let us now focus on the spin-tidal corrections coming from the ordinary TLNs.
Their PN order can be computed again by noticing that $Q^L$ (respectively, $S^L$) enters the waveform at $l$PN (respectively,
$(l+1/2)$PN) order, as discussed above. On the other hand, the leading-order spin terms in $G^{L}$ and $H^{L}$
enter, respectively, at $(l+1+3/2)$PN and at $(l+3/2+1/2)$PN order. Therefore, the overall, leading-order, spin-tidal
contributions of the $\sim Q^L G^L$ and $\sim S^L H^L$ couplings both enter at $(2l+5/2)$PN order.

To summarize, the leading-order, spin-tidal corrections coming from the excitation of $l$-pole moments at linear order in
the spin read
\begin{align}
 {\rm PN~order}_{\rm spin-TLNs}=&\,2l+\frac{5}{2}\,,\nonumber\\
 {\rm PN~order}_{\rm RTLNs}=&\,2l+\frac{1}{2}+2\delta_{l2}\,,
\end{align}
where the first and second line refer to terms proportional to the ordinary TLNs and to the RTLNs, respectively.
Interestingly, the PN orders of the two contributions coincide only when $l=2$, yielding the $6.5$PN terms discussed in
this work [the terms $\tilde\Lambda$, $\tilde\Sigma$, and $\tilde\Gamma$ in Eq.~\eqref{PHASE}]. For $l\geq3$, the
contribution from the RTLNs is always dominant.

\subsection{Are Lagrangian formulation and perturbation theory compatible?} \label{sec:issue}
In Ref.~\cite{Pani:2015nua}, \emph{four} RTLNs were introduced to describe (at linear order in the spin) the coupling
between $l=2,3$ multipole moments of a spinning object with $l=2,3$ tidal moments. According to the selection rules
described in Ref.~\cite{Pani:2015nua}, $\lambda_{23}$ describes how a mass quadrupole moment is induced by an octupolar
magnetic tidal moment at linear order in the spin, whereas $\sigma_{32}$ describes how a current octupole moment is
induced by a quadrupolar tidal moment. A similar argument applies to $\lambda_{32}$ and $\sigma_{23}$.

However, as previously discussed, our interaction Lagrangian~\eqref{L2int} contains only \emph{two} coupling terms
proportional to the spin and which are responsible for the coupling between multipole moments and tidal moments with
opposite parity and $l\leftrightarrow l\pm1$. In other words, a Lagrangian formulation seems to predict \emph{two}
RTLNs, rather than the four RTLNs that have been explicitly computed in Ref.~\cite{Pani:2015nua}.

One might be tempted to think that a relation exists between $\lambda_{23}$ and $\sigma_{32}$ (and between
$\lambda_{32}$ and $\sigma_{23}$) so that, once the four RTLNs are explicitly computed, they would satisfy the
relations~\eqref{rotTLNs}. Unfortunately, we have checked if this is the case by explicitly computing the RTLNs for
neutron stars using perturbation theory as discussed in Ref.~\cite{Pani:2015nua}, and found no numerical evidence for a
relation between those RTLNs.
In fact, we believe that such putative relation can hardly emerge from the perturbed Einstein equations, since
electric-led and magnetic-led RTLNs belong to two different sectors, namely to Zerilli and Regge-Wheeler perturbations,
respectively. While it is true that the two sectors enjoy some special symmetries in the case of Schwarzschild black
holes~\cite{Chandra}, such symmetries are broken for material bodies and we do not see any reason why the corresponding
RTLNs should be related by (truly) universal relations which should be completely independent of the body
composition\footnote{We stress that, although there is some tension between some of the RTLNs computed in
  Ref.~\cite{Pani:2015nua} and those computed by other groups~\cite{Gagnon-Bischoff:2017tnz}, the fact that the
  electric-led and magnetic-led RTLNs are independent should not be affected by such discrepancy.}.

On the other hand, the fact that in the approach presented here only two RTLNs are independent seems intrinsically
related with the Lagrangian formulation, which clearly introduces the same coupling constant in two different
Euler-Lagrange equations. To better illustrate this point, let us make a specific example. The coupling
\begin{align}
  \mathcal{L}_2^{int} &\supset \alpha J_2^a Q^{bc} S^{abc}
\end{align}
contributes to the Euler-Lagrange equations for both $Q^{ab}$ and $S^{abc}$. In the former case, it gives a term $\sim \alpha J_2^a
S^{abc}$, whereas in the latter case it gives $\sim\alpha J_2^a Q^{bc}$. In both cases, the terms depend on the same
coupling factor, $\alpha$.

Unfortunately, at the moment we are not able to explain this apparent inconsistency. One option could be that a
Lagrangian formulation fails to reproduce the full couplings that arise in perturbation theory; however, we consider
this option as unlikely. Other possible explanations could come from a nontrivial static limit of the dynamical action
describing the time evolution of the induced multipole moments~\cite{Steinhoff:2016rfi}, or by the role of the internal
fluid dynamics, or finally by some hidden symmetry of the perturbation equations that effectively reduces the number of
independent RTLNs to two. We plan to investigate this issue elsewhere. We stress, however, that the expression for the
GW phase in Eq.~\eqref{PHASE} can also accommodate putative relations among the RTLNs.

\section{Discussion and outlook}
We have computed, for the first time, the spin-tidal couplings that modify the dynamics of two orbiting bodies in
general relativity at the leading PN order and at linear order in the spin. These corrections depend on both the standard TLNs and on the RTLNs recently introduced
in previous work.  Our main result is Eq.~\eqref{PHASE}, which provides the new spin-tidal terms for the GW phase of
circular binaries with spins orthogonal to the orbital plane. All these new terms modify the phase at $1.5$PN order
relative to the standard, quadrupolar, tidal deformability term at the leading order. At linear order in the spin, the
terms computed here should include \emph{all} the tidal terms up to $6.5$PN order.
The new terms computed here enter the GW phase at a lower order relative to the standard, octupolar tidal terms. We
proved that this is the case for any RTLN with $l\geq3$.

We have encountered a conceptual problem related to the inclusion of the RTLNs in the Lagrangian formulation. We hope that
our results will motivate more work which may shed light on this issue.

An analysis of the impact of spin-tidal couplings in the parameter estimation of binary NSs is ongoing and will
appear in a follow-up paper~\cite{Jimenez-Forteza:2018buh}.

Another application of our results is related to GW searches for exotic compact
objects~\cite{Cardoso:2017cqb,Cardoso:2017njb}. Since the TLNs of a black hole are
zero~\cite{Binnington:2009bb,Damour:2009vw,Damour:2009va,Porto:2016zng}, measuring the effect of the tidal deformability
in the waveform of a binary coalescence provides an independent way to distinguish black holes from other exotic compact
alternatives~\cite{Cardoso:2017cfl,Sennett:2017etc,Maselli:2017cmm,Johnson-McDaniel2018}. There is no reason to expect
that black-hole mimickers should be slowly spinning (this is particularly true for supermassive objects in the LISA
band, whose spin might grow through accretion). Thus, the inclusion of the spin-tidal couplings computed here will
greatly improve previous analysis~\cite{Maselli:2017cmm}.

{\it Note added}. -- After completion of this work, we had been informed of a related work by
Landry~\cite{Landry:2018bil}. Beside the different notation, our work differs from Ref.~\cite{Landry:2018bil} because it
includes also the spin-tidal terms proportional to the ordinary TLNs. While our result for the energy flux agrees with
that of Ref.~\cite{Landry:2018bil} in the appropriate particular case, our result for the GW phase does not agree with
that derived in Ref.~\cite{Landry:2018bil}. We believe that the source of discrepancy is a different definition of the
energy of the binary system. We also note that our results for both the energy flux and the GW phase agree with those of
Refs.~\cite{Yagi:2013sva,Banihashemi:2018xfb} when neglecting spin effects.

\begin{acknowledgments}
We are indebted to Eanna Flanagan, Philippe Landry, Eric Poisson, and Jan Steihoff for comments on a preliminary draft, to Justin Vines for useful suggestions, and to Tanja Hinderer and Kent Yagi for interesting conversation during the development of this project. We are also grateful to Kent Yagi and Justin Vines for
comparisons related to the magnetic tidal Love numbers in the nonspinning case, and to Ira Rothstein and Leo Stein for further discussion.
P.P. acknowledges financial support provided under the European Union's H2020 ERC, Starting Grant Agreement
No.~DarkGRA--757480.
This project has received funding from the European Union's Horizon 2020 research and innovation program under the
Marie Sklodowska-Curie Grant Agreement No. 690904.
The authors would like to acknowledge networking support by the COST Action CA16104.
\end{acknowledgments}
\ \\

\appendix
\section{Multipole moments transformation between body frame and global frame}\label{app:momtransf}
The global multipole moments of a system of $N$ structured bodies, $M^L_{g,A}$, $Z^{iL}_{g,A}$, can be expressed in
terms of the body-frame mass and current multipole moments $M^L_{A}$, $J^{L}_{A}$ (up to $1$PN order for the mass
moments, $0$PN order for the current moments) as~\cite{Racine:2004hs,Vines:2013prd}:
\begin{align}
  M^L_{g,A} =&  M^L_A +c^{-2} \bigg[\left(\frac{3}{2} v_A^2 -(l+1) G_{g,A} \right)M^L_A\nonumber\\
    &-\frac{2l^2+5l-5}{(l+1)(2l+3)} v^j_A \dot{M}^{jL}_A\nonumber\\
  &  -\frac{2l^3+7l^2+16l+7}{(l+1)(2l+3)} a^j_A M^{jL}_A \nonumber\\
    & -\frac{2l^2+17l-8}{2(2l+1)} v^{j \langle a_l}_A M^{L-1 \rangle j}_A \nonumber\\
    &+\frac{4l}{l+1} v^j_A \epsilon^{jk \langle a_l} J^{L-1 \rangle k}_A \bigg]
  + O \left(c^{-4} \right)\,,\label{eq:Mg}\\
  Z^{iL}_{g,A} =&  \frac{4}{l+1} \dot{M}^{iL}_A +4 v^i_A M^L_A\nonumber\\
  &-\frac{4(2l-1)}{2l+1} v^j_A M^{j \langle L-1}_A
  \delta^{a_l \rangle i} \nonumber\\
  &-\frac{4l}{l+1} \epsilon^{ji \langle a_l} J^{L-1 \rangle j}_A
  + O \left(c^{-2} \right)\,,\label{eq:Zg}
\end{align}
where
\begin{equation}
\label{eq:globalG0}
G_{g,A} =  \sum_{B \neq A} \sum_{k=0}^{\infty} \frac{(-1)^k}{k!} M_B^K
\partial^{(A)}_{K} \frac{1}{|z_A-z_B|} + O \left( c^{-2}\right)\,.
\end{equation}

The tidal moments in the body frame $G_A^L(s_A)$, $H_A^L(s_A)$, which enter in Eq.~\eqref{eq:dipole} and then
in the orbital equations of motion, can be expressed in terms of the global multipole moments as follows.
At Newtonian order,
\begin{align}
  G^i_{A} = & G_{g,A}^i- {\ddot z}^i_A+O(c^{-2})\,,\label{eq:Ga}\\
  G^L_{A} = & G_{g,A}^L+O(c^{-2}) \qquad l \geq2\,,\label{eq:G0}\\
H^L_A =& Y_{g,A}^{jk \langle L-1} \epsilon^{a_l \rangle jk}-4v_A^jG_{g,A}^{k\langle L-1}\epsilon^{a_l\rangle jk} \nonumber \\
& - l! \Lambda_{\zeta,A}^{jk \langle L-1} \epsilon^{a_l\rangle jk}  + O(c^{-2})  \quad l  \geq 1 \label{eq:H}\,,
\end{align}
where the global-frame electric and magnetic tidal moments are
\begin{align}
G^L_{g,A} =&  \sum_{B \neq A} \sum_{k=0}^{\infty} \frac{(-1)^k}{k!} M_B^K
\partial^{(A)}_{KL} \frac{1}{|z_A-z_B|} + O \left( c^{-2}\right)\label{eq:Gg}\,,\\
Y^{iL}_{g,A} =& \sum_{B \neq A} \sum_{k=0}^{\infty} \frac{(-1)^k}{k!} Z^{iK}_{g,B}\,
\partial^{(A)}_{KL} \frac{1}{|z_A-z_B| } + O \left(c^{-2} \right)\,,\label{eq:Yg}
\end{align}
and $Z^{iK}_{g,B}$ is given in Eq.~\eqref{eq:Zg}.

\begin{widetext}
At $1$PN order, the electric tidal moments are
  \begin{align}
  G^{L}_{A} = & F_{g,A}^L- l! \Lambda^L_{\Phi,A} +\frac{1}{c^2}
  \bigg[ \dot{Y}^{\langle L \rangle}_{g,A}-v^j_A Y^{jL}_{g,A}
  +(2v^2_A-lG_{g,A})G^L_{g,A} -(l/2)v^{j\langle a_l}_A G^{L-1 \rangle j}_{g,A} +
  (l-4) v^{\langle a_l}_A \dot{G}^{L-1 \rangle }_{g,A}\nonumber\\
  &-(l^2-l+4) a^{\langle a_L}_A G^{L-1 \rangle}_{g,A}
  -(l-1)! \dot{\Lambda}_{\zeta,A}^{\langle L \rangle} \bigg] + O \left(c^{-4} \right) \qquad l \geq 1 \,, \label{eq:G}
  \end{align}
where
\begin{align}
  F^{L}_{g,A} = &\sum_{B \neq A} \sum_{k=0}^{\infty} \frac{(-1)^k}{k!}
  \bigg[ N^K_{g,B}\, \partial^{(A)}_{KL} \frac{1}{|z_A-z_B| }
    +\frac{1}{2c^2} P^K_{g,B} \, \partial^{(A)}_{K\langle L \rangle}
    |z_A-z_B|\bigg]+ O \left(c^{-4} \right)\label{eq:Fg}\\
  N^{L}_{g,A} = &M^L_{g,A} + \frac{1}{(2l+3)c^2} [v^2_A M^L_A +2 v^j_A \dot{M}^{jL}_A
    +2lv^{j\langle a_l}_A M^{L-1 \rangle j}_A + a^j_A M^{jL}_A ] + O \left(c^{-4} \right)\label{eq:Ng}\\
  P^{L}_{g,A} =&   \ddot{M}^L_{A} +2 l v_A^{\langle a_l} \dot{M}_A^{L-1 \rangle}
  +l a_A^{\langle a_l} M_A^{L-1 \rangle}
  + l(l-1)v_A^{\langle a_l a_{l-1}} M_A^{L-2\rangle} + O \left(c^{-2} \right)\label{eq:Pg}\\
\Lambda^i_{\Phi,A} = & a^i_A + \frac{1}{c^2} \bigg[(v^2_A+G_{g,A})a^i_A+\frac{1}{2} v^{ij}_A a_A^j
  +2 \dot{G}_{g,A} v^i_A \bigg]  + O \left(c^{-4} \right)\nonumber\\
\Lambda^{ij}_{\Phi,A} = &  \frac{1}{c^2} \left(-\frac{1}{2} a_A^{\langle ij \rangle} + v_A^{\langle i }
\dot{a}_A^{j \rangle} \right)+ O \left(c^{-4} \right) \nonumber\\
\Lambda^{L}_{\Phi,A} = & \, 0 \qquad l \geq 3\label{eq:Lambdaph}\\
\end{align}
\end{widetext}
\begin{align}
\Lambda^i_{\zeta,A} = & -2 G_{g,A} v^i_A  + O \left(c^{-2} \right)\nonumber\\
\Lambda^{ij}_{\zeta,A} = & -\frac{3}{2} v^{[i}_A a^{j]}_A -2 v^{\langle i}_A
a_A^{j \rangle}-\frac{4}{3} \dot{G}_{g,A} \delta^{ij} + O \left(c^{-2} \right)\nonumber\\
\Lambda^{ijk}_{\zeta,A} = & -\frac{6}{5} \delta^{i \langle j}
\dot{a}_A^{k \rangle} + O \left(c^{-2} \right)\nonumber \\
\Lambda^{L}_{\zeta,A} = & \, 0 \qquad l \geq 4\label{eq:Lambdaz}\,.
\end{align}
%
\section{Tidal moments in the truncated system}\label{app:tidalpot}
We here compute the tidal moments for the system considered in this paper in which the body $1$ has nonvanishing mass
$M_1$ and spin $J^a_1$, the body $2$ has nonvanishing mass $M_2$, spin $J^a_2$, quadrupole moments $Q^{ab}=M_2^{ab}$,
$S^{ab}=J_2^{ab}$ and octupole moments $Q^{abc}=M_2^{abc}$, $S^{abc}=J_2^{abc}$. We only focus on the tidal moments of the
body $2$, $G^L_2$, $H^L_2$, with $l=2,3$, because these are those which induce the multipole moments of our truncation
in the adiabatic relations~\eqref{eq:adiabatic}. The other tidal moments needed to derive the orbital equations of
motion (i.e., $G_1^i$, $H_1^{ij}$, $G_2^L$, and $H_2^L$ with $l=1$ and $l=4$) can be obtained in a similar way.  As
discussed in Sec.~\ref{subsec:eom}, we compute the electric, quadrupolar tidal moment $G_2^{ab}$ at $1$PN order,
while all the other tidal moments are computed at $0$PN order only. These are the contributions needed in order to
compute the waveform at $6.5$PN order (see Sec.~\ref{subsec:waveform}).

At $0$PN, Eq.~\eqref{eq:Gg} gives $G^L_{g,2}=M_1\partial_L\frac{1}{r}+O(c^{-2})$, i.e.,
\begin{align}
G^{ab}_{g,2}=&\frac{3M_1}{r^3}n^{\langle ab\rangle}+O(c^{-2}) \label{eq:Ggl23a}\\
G^{abc}_{g,2}=&-\frac{15M_1}{r^4}n^{\langle abc\rangle}+O(c^{-2})\,,\label{eq:Ggl23b}
\end{align}
and, since [see Eq.~\eqref{eq:Zg}]
\begin{align}
  Z^i_{g,1}=&4M_1v^i_1+O(c^{-2})\nonumber\\
  Z^{ij}_{g,1}=&-2\epsilon^{ijk}J_1^k+O(c^{-2})\nonumber\\
  Z^L_{g,1}=&O(c^{-2})\qquad\qquad(l\ge3)\,,
\end{align}
Eq.~\eqref{eq:Yg} gives
\begin{equation}
Y^{iL}_{g,2}=4M_1v_1^i\partial_L\frac{1}{r}+2\epsilon^{ijk}J_1^k\partial_{jL}\frac{1}{r}+O(c^{-2})\,.
\end{equation}
Replacing in Eqs.~\eqref{eq:G0}, \eqref{eq:H}, we find (since $v^i=v_2^i-v_1^i$)
\begin{align}
  G_2^L=&M_1\partial_L\frac{1}{r}+O(c^{-2})  \qquad\qquad(l\ge2)\nonumber\\
  H_2^L=&-4M_1v^i\partial_{c\langle L-1}\frac{1}{r}\epsilon^{a_l\rangle ic}
  +2\epsilon^{ijk}J_1^k\partial_{jc\langle L-1}\frac{1}{r}\epsilon^{a_l\rangle ic}\nonumber\\
  &+O(c^{-2})\qquad\qquad(l\ge2)\,.
\end{align}
Therefore, since $\partial_L\frac{1}{r}=(-1)^l(2l-1)!!\frac{n^{\langle L\rangle}}{r^{l+1}}$,
\begin{align}
G_2^{ab}=&\frac{3\eta_1 M}{r^3}n^{\langle ab\rangle}+O(c^{-2})\,,\label{eq:G2ab}\\
G_2^{abc}=&-\frac{15\eta_1M}{r^4}n^{\langle abc\rangle}+O(c^{-2})\,,\label{eq:G2abc}
\end{align}
\begin{widetext}
\begin{align}
H_2^{ab}=&\frac{6\eta_1M}{r^3}v^d\left(n^{ac}\epsilon^{bcd}+n^{bc}\epsilon^{acd}\right)+\frac{30J_1^c}{r^4}n^{\langle abc\rangle}+O(c^{-2})\,,\label{eq:H2ab}\\
H_2^{abc}=&-\frac{20\eta_1M}{r^4}v^e\left(n^{\langle dab\rangle}\epsilon^{cde}+n^{\langle dbc\rangle}\epsilon^{ade}+n^{\langle dca\rangle}\epsilon^{bde}\right)-\frac{210 J_1^d}{r^5}n^{\langle abcd\rangle}+O(c^{-2})\,.\label{eq:H2abc}
\end{align}
At $1$PN order, Eqs.~\eqref{eq:G}--\eqref{eq:Lambdaz} give
\begin{align}
  G_2^{ab}=&\frac{3\eta_1M}{r^3}n^{\langle ab\rangle}
  +\frac{1}{c^2}\frac{3\eta_1M}{r^3}\left[\left(2v^2-\frac{5\eta^2_2}{2}
    {\dot r}^2\right.\right.-\frac{5+\eta_1}{2}\frac{M}{r}\bigg)n^{\langle ab\rangle}+v^{\langle ab\rangle}-(3-\eta_2^2){\dot r}n^{\langle a}v^{b\rangle}\bigg] \nonumber \\
& + \frac{6  }{c^2 r^4}  J_1^d v^e \epsilon^{ec \langle a} \left( 5 n^{b \rangle cd}- \delta^{b \rangle d} n^c- n^{b \rangle }\delta^{cd} \right)   +O(c^{-4})\,.\label{eq:G21PN}
\end{align}

\section{Orbital equations of motion of the two-body system}\label{app:eom_tr}
We here show the explicit expression of the orbital equations of motion given in Eqs.~\eqref{eq:orbeom12a} and~\eqref{eq:orbeom12b},
\begin{align}
  M_1  a^i_1 = & F^i_{1,M} +  F^i_{1,J} + F^i_{1,Q2}+ F^i_{1,Q3} +  F^i_{1,S2}  +  F^i_{1,S3} \,,\\
    M_2  a^i_2 = & F^i_{2,M} +  F^i_{2,J} + F^i_{2,Q2}+ F^i_{2,Q3} +  F^i_{2,S2}  +  F^i_{2,S3} \,.
\end{align}
The mass monopole contributions are
\begin{align}
  F^i_{1,M} = & \frac{M_1 M_2}{r^2} n^i + \frac{1}{c^2}
  \frac{M_1 M_2}{r^2} \left\{n^i \left[ 2 v^2 -v_1^2
    -\frac{3}{2} \left(n^a v^a_2\right)^2-\frac{5 M_1}{r}
    -\frac{4 M_2}{r} \right]+ v^i n^a \left(4 v^a_1-3 v^a_2\right) \right\} +O(c^{-4})\,,\\
  F^i_{2,M} = & - \frac{M_1 M_2}{r^2} n^i - \frac{1}{c^2}
  \frac{M_1 M_2}{r^2} \left\{n^i \left[ 2 v^2 -v_2^2
    -\frac{3}{2} \left(n^a v^a_1\right)^2-\frac{4 M_1}{r} -\frac{5 M_2}{r} \right]
  - v^i n^a \left(4 v^a_2-3 v^a_1\right) \right\}+O(c^{-4})\,.
\end{align}
The spin contributions are
\begin{align}
  F^i_{1,J} = & \frac{1}{c^2} \frac{M_1}{r^3} \epsilon^{abc} J_2^c \left[\delta^{ai}
    \left(4 v^b-6n^{bd} v^d \right) -6 n^{ai} v^{b}\right]
  -\frac{1}{c^2} \frac{M_2}{r^3} \epsilon^{abc} J_1^c \left[3\delta^{ai} \left(n^{bd} v^d -v^b\right) +6 n^{ai} v^{b}\right]
  +O(c^{-4}) \,, \\
  F^i_{2,J} = & \frac{1}{c^2}\frac{M_1}{r^3} \epsilon^{abc} J_2^c
  \left[3\delta^{ai} \left(n^{bd} v^d -v^b\right) +6 n^{ai} v^{b}\right]
  -\frac{1}{c^2} \frac{M_2}{r^3} \epsilon^{abc} J_1^c \left[\delta^{ai} \left(4 v^b-6n^{bd} v^d \right) -6 n^{ai} v^{b}\right]
  +O(c^{-4})\,.
\end{align}
The mass quadrupole contributions are
\begin{align}
  F^i_{1,Q2} &=  \frac{3 M_1}{2 r^4} Q^{ab} \left(5n^{abi}-2n^a \delta^{bi} \right)
  + \frac{1}{c^2} \bigg( \frac{3 M_1}{2 r^4} Q^{ab} \bigg\{ 5 n^{abi} \left[2 v^2-v_1^2 -
    \frac{7}{2} ( n^c v_2^c)^2 \right.\nonumber\\
    &\left.-\frac{47 M_1}{5 r} -\frac{24 M_2}{5 r} \right]
  -2 n^a \delta^{bi} \left[2 v^2-v_1^2 -\frac{5}{2} \left(n^c v_2^c \right)^2 -\frac{19 M_1}{2 r}
    -\frac{4 M_2}{r} \right] + n^a v_2^{bi}  \nonumber\\
  &+ (5 n^{ai}- \delta^{ai} ) v_2^{bc} n^c 
   + v^i (5 n^{abc}-2n^a \delta^{bc} ) (4 v_1^c -3 v_2^c ) \bigg\} + \frac{3 M_1}{2 r^3} \dot{Q}^{ab}
  [ n^{ab} (5 v_2^c\, n^{ci} + 3v^i ) -4 v^a n^{bi} \nonumber\\
    & - 2 \delta^{ai} n^{bc} (2v_1^c-v_2^c ) ] -\frac{3 M_1}{4 r^2}
  \ddot{Q}^{ab}( n^{abi}+2 n^a \delta^{bi} ) \bigg)\nonumber\\
  & - \frac{3}{c^2} \epsilon^{icd} J_1^c \bigg\{ \frac{Q^{ab}}{r^5} \bigg[ \frac{5}{2} n^{ab} (7 n^{de} v^e
    - v^d) +(\delta^{ad} -5 n^{ad}) v^b -5 \delta^{ad} n^{be} v^e \bigg] + \frac{\dot{Q}^{ab}}{r^4}
  (\delta^{ad}-\frac{5}{2}n^{ad}) n^b \bigg\} \nonumber\\
  & - \frac{3 \epsilon^{cde} J_1^a}{c^2} \bigg\{ \frac{Q^{bd}}{r^5} \bigg[ 5 n^{ab}
    \left( \delta^{ic} v^e - \delta^{ie} v^c \right) +
    5 n^{ac} \left( \delta^{ib} v^e - \delta^{ie} v^b\right) + 5 n^{bc}
    \left( \delta^{ia} v^e - \delta^{ie} v^a \right) \nonumber\\
    & + 35 n^{abc} \left( \delta^{ie} n^f v^f -n^i \right) + \delta^{ab}
    \left(\delta^{ie} v^c -\delta^{ic}  \right) + \delta^{ac}
    \left(\delta^{ie} v^b -\delta^{ib}  \right)  \nonumber\\
    & + 5 \left(\delta^{ab} n^c + \delta^{ac} n^b \right) \left(n^i -
    \delta^{ie} n^f v^f \right) \bigg] - \frac{ \dot{Q}^{bd}}{r^4} \delta^{ie}
  \left(5 n^{abc}-\delta^{ab} n^c -\delta^{ac} n^b \right) \bigg\} +O(c^{-4})\,,
\end{align}
\begin{align}
  F^i_{2,Q2} = &  - \frac{3 M_1}{2 r^4} Q^{ab} \left(5n^{abi}-2n^a \delta^{bi} \right)
  + c^{-2} \bigg( \frac{3 M_1}{2 r^4} Q^{ab} \bigg\{ -5 n^{abi} \left[2 v^2-v_2^2 -
    \frac{7}{2} ( n^c v_1^c)^2 -\frac{8 M_1}{r} -\frac{6 M_2}{r} \right] \nonumber\\
  & +2 n^a \delta^{bi} \left[3 v^2-v_2^2 -5 \left(n^c v^c \right)^2
    -\frac{5}{2} \left(n^c v_1^c \right)^2 -\frac{8 M_1}{r} -\frac{11 M_2}{2r} \right]
  + n^i v^{ab} + 5 n^{aci}(2v^b v_1^c-  v_2^{bc}) \nonumber\\
  & + v^i (5 n^{abc}-2n^a \delta^{bc} ) (4 v_2^c -3 v_1^c ) + n^a v_2^b (v_2^i-2 v_1^i)
  + \delta^{bi} n^c[(5 v_2^a-4v_1^a)v_2^c -6 v^a v_1^c]\bigg\} \nonumber\\
  & + \frac{3 M_1}{r^3} \dot{Q}^{ab} [ v^b (2 n^{ai}-\delta^{ai}) +\delta^{ai} n^{bc} v^c
    - 2 n^{ab} v^i ] \bigg) \nonumber\\
  & + \frac{3 M_1}{c^2M_2} \epsilon^{icd} J_2^c \bigg\{ \frac{Q^{ab}}{r^5}
  \bigg[ \frac{5}{2} n^{ab} (7 n^{de} v^e - v^d) +(\delta^{ad} -5 n^{ad}) v^b
    -5 \delta^{ad} n^{be} v^e \bigg] + \frac{\dot{Q}^{ab}}{r^4} (\delta^{ad}-
  \frac{5}{2}n^{ad}) n^b \bigg\} \nonumber\\
  & + \frac{3 \epsilon^{cde} J_1^a}{c^2} \bigg\{ \frac{Q^{bd}}{r^5}
  \bigg[ 5 n^{ab} \left( \delta^{ic} v^e - \delta^{ie} v^c \right) +
    5 n^{ac} \left( \delta^{ib} v^e - \delta^{ie} v^b\right) + 5 n^{bc} \left( \delta^{ia} v^e
    - \delta^{ie} v^a \right) \nonumber\\
    & + 35 n^{abc} \left( \delta^{ie} n^f v^f -n^i \right) + \delta^{ab} \left(\delta^{ie} v^c
    -\delta^{ic}  \right) + \delta^{ac} \left(\delta^{ie} v^b -\delta^{ib}  \right)  \nonumber\\
    & + 5 \left(\delta^{ab} n^c + \delta^{ac} n^b \right) \left(n^i -\delta^{ie} n^f v^f \right) \bigg]
  - \frac{ \dot{Q}^{bd}}{r^4} \delta^{ie} \left(5 n^{abc}-\delta^{ab} n^c -\delta^{ac} n^b \right) \bigg\}+O(c^{-4})\,.
\end{align}
The mass octupole contributions are
\begin{align}
F^i_{1,Q3} = & - \frac{5 M_1}{2r^5} Q^{abc} \left( 7 n^{iabc} - 3 \delta^{ic} n^{ab} \right) +O(c^{-2})\,,\\
F^i_{2,Q3} = &  \frac{5 M_1}{2r^5} Q^{abc} \left( 7 n^{iabc} - 3 \delta^{ic} n^{ab} \right)+O(c^{-2})\,.
\end{align}
The current quadrupole contributions are
\begin{align}
  F^i_{1,S2} = & - \frac{4 M_1 \epsilon^{bcd}}{c^2 }
  \bigg\{ \frac{S^{ad}}{r^4} \bigg[ n^a \left(\delta^{ib} v^c -\delta^{ic} v^b \right) +
n^b \left(\delta^{ia} v^c -\delta^{ic} v^a \right) + 5 n^{ab} \left(\delta^{ic} n^e v^e -n^i v^c \right) \bigg] \nonumber\\
& - \frac{ \dot{S}^{ad}}{r^3} \delta^{ic} n^{ab}  \bigg\}\nonumber \\
  & - \frac{ J_1^c}{c^2} \frac{S^{ab} }{r^5} \left[ 4 \delta^{bc}
    \left( 5 n^{ia} - \delta^{ia} \right) + 10 \left( \delta^{ia} n^{bc} + \delta^{ib} n^{ac} + 
\delta^{ic} n^{ab}   \right)- 70 n^{iabc} \right]+O(c^{-4})\,,\\
  F^i_{2,S2} = & \frac{4 M_1 \epsilon^{bcd}}{c^2 } \bigg\{ \frac{S^{ad}}{r^4}
  \bigg[ n^a \left(\delta^{ib} v^c -\delta^{ic} v^b \right) +
n^b \left(\delta^{ia} v^c -\delta^{ic} v^a \right) + 5 n^{ab} \left(\delta^{ic} n^e v^e -n^i v^c \right) \bigg] \nonumber\\
& - \frac{ \dot{S}^{ad}}{r^3} \delta^{ic} n^{ab}  \bigg\} \nonumber\\
  & + \frac{ J_1^c}{c^2} \frac{S^{ab}}{r^5} \left[ 4 \delta^{bc}
    \left( 5 n^{ia} - \delta^{ia} \right) + 10 \left( \delta^{ia} n^{bc} + \delta^{ib} n^{ac} + 
  \delta^{ic} n^{ab}   \right)- 70 n^{iabc} \right]+O(c^{-4})\,.
\end{align}
The current octupole contributions are
\begin{align}
  F^i_{1,S3} = & - \frac{15 M_1}{c^2} \bigg\{ \frac{S^{bde}}{r^5}  n^{ab}  v^c
  \bigg[\frac{7}{2}(\epsilon^{iae} n^{c}-\epsilon^{cae} n^{i})n^d
    -( \epsilon^{iad}\delta^{ce} + \epsilon^{icd}\delta^{ae} +
    \epsilon^{acd}\delta^{ie} )\bigg] - \frac{\dot{S}^{bde}}{2r^4} \epsilon^{iad} n^{abe} \bigg\}\nonumber\\
  & - \frac{45 J_1^c}{4 c^2} \frac{S^{abc}}{r^6} \bigg[ \delta^{cd}
    \left(\delta^{ia} n^b + \delta^{ib} n^a -7 n^{iab} \right)
    - \frac{7}{3} \left( \delta^{ia} n^{bcd} +\delta^{ib} n^{acd} +
    \delta^{ic} n^{abd} + \delta^{id} n^{abc} -9 n^{iabcd} \right) \bigg]\nonumber\\
  &+O(c^{-4})\,,\\
  F^i_{2,S3} = &  \frac{15 M_1}{c^2} \bigg\{ \frac{S^{bde}}{r^5}  n^{ab}  v^c
  \bigg[\frac{7}{2}(\epsilon^{iae} n^{c}-\epsilon^{cae} n^{i})n^d
    -( \epsilon^{iad}\delta^{ce} + \epsilon^{icd}\delta^{ae} +
    \epsilon^{acd}\delta^{ie} )\bigg] - \frac{\dot{S}^{bde}}{2r^4} \epsilon^{iad} n^{abe} \bigg\} \nonumber\\
  & + \frac{45 J_1^c}{4 c^2} \frac{S^{abc}}{r^6} \bigg[ \delta^{cd}
    \left(\delta^{ia} n^b + \delta^{ib} n^a -7 n^{iab} \right)
    - \frac{7}{3} \left( \delta^{ia} n^{bcd} +\delta^{ib} n^{acd} +
    \delta^{ic} n^{abd} + \delta^{id} n^{abc} -9 n^{iabcd} \right) \bigg]\nonumber\\
  &+O(c^{-4})\,.
\end{align}
\ \\

\end{widetext}

\section{Higher-order terms in the GW phase}\label{app:higher}
For completeness, we provide the higher-order terms entering the GW phase~\eqref{PHASE} (i.e., appearing at $7$PN order
and beyond) which are proportional to the TLNs. These terms are computed as a by-product of our analysis and could be
useful for comparison. The extra terms in Eq.~\eqref{PHASE} read
 \begin{align}
   \psi(x)\supset&\left(\frac{4000}{9} - \frac{4000}{9\eta_1} \right)
   \frac{\lambda_3^{(1)}}{M^7}   x^7 \nonumber\\
   &+\left( \frac{29400}{11}  -
   \frac{29400}{11 \eta_1} \right) \frac{\sigma_3^{(1)}}{M^7}  x^8 \nonumber\\
  & + \left( -\frac{44800}{3}  + \frac{22400}{3 \eta_1}
    + \frac{22400 \eta_1}{3 } \right)  \chi_2  \frac{\sigma_3^{(1)}}{M^7}  x^{8.5} \nonumber\\
  & +(1\leftrightarrow 2)\,.  
 \end{align}
To the best of our knowledge, some of these terms have never been published before. 

\section{Comparison to the RTLNs defined in Ref.~\cite{Pani:2015nua}}\label{app:comparison}
We here show the relations between the TLNs defined in Eq.~\eqref{eq:adiabatic} and those of Ref.~\cite{Pani:2015nua}. In the following, we set the speed of light $c=1$.

In Ref.~\cite{Pani:2015nua}, the standard electric and magnetic TLNs are defined as
\begin{equation}
 \lambda_E^{(l)}\equiv \frac{\partial M_l}{\partial {\cal E}_{l}}\,,\qquad 
 \lambda_M^{(l)}\equiv \frac{\partial S_l}{\partial {\cal B}_{l}}\,, \label{Love0}
\end{equation}
where the multipole moments $M_l$, $S_l$, and the tidal-field components ${\cal E}_{l}$, ${\cal B}_{l}$ are given in
terms of the asymptotic expansion of the metric 
(see Refs.~\cite{Binnington:2009bb,Cardoso:2016ryw} and Appendix~B of Ref.~\cite{Cardoso:2017cfl}).
With the above definitions, $\lambda_l$ and $\sigma_l$ used in this work are, respectively, given by
\begin{align}
\lambda^{(l)}_E =& -\frac{(2l-1)!!}{l(l-1)} \sqrt{\frac{2l+1}{4\pi}} \lambda_l  \,, \\
\lambda^{(l)}_M =& -\frac{4(2l-1)!!}{3(l-1)} \sqrt{\frac{2l+1}{4\pi}} \sigma_l  \,,
\end{align}
with $l \geq 2$.

On the other hand, the RTLNs in the axisymmetric case are defined in Ref.~\cite{Pani:2015nua} as
\begin{equation}
 \delta\lambda_{E}^{(ll')}\equiv \frac{\partial M_l}{\partial {\cal B}_{l'}}\,,\qquad 
 \delta\lambda_{M}^{(ll')}\equiv \frac{\partial S_l}{\partial {\cal E}_{l'}}\,,
\end{equation}
with the same moments given in 
Refs.~\cite{Binnington:2009bb,Cardoso:2016ryw,Pani:2015nua,Cardoso:2017cfl}.
With the above definitions, the RTLNs $\lambda_{23}$, $\sigma_{23}$, $\lambda_{32}$ and $\sigma_{32}$ defined here are
related to those defined in Ref.~\cite{Pani:2015nua} by the relations
\begin{align}
\delta\lambda^{(23)}_E =& -\sqrt{\frac{7}{\pi}} J \lambda_{23} \nonumber \,, \\
\delta\lambda^{(23)}_M =& -\sqrt{\frac{7}{\pi}} J \sigma_{23} \nonumber \,, \\
\delta\lambda^{(32)}_E =& -\frac{15}{\sqrt{5\pi}} J \lambda_{32} \nonumber \,, \\
\delta\lambda^{(32)}_M =& -\frac{45}{8\sqrt{5\pi}} J  \sigma_{32}  \,,
\end{align}
where $J$ is the absolute value of the angular momentum of the body which is deformed.

\bibliography{refs}
\end{document}